\documentclass[prd,aps,nofootinbib,11pt]{revtex4}
\usepackage{amssymb}
\usepackage{amsmath,graphicx,color,epsfig}

\begin{document}
\title{$B\to S$ Transition Form Factors in the PQCD approach}
\author{Run-Hui Li$^{1,2}$, Cai-Dian L\"{u}$^{2,3}$, Wei Wang$^{2}$  and Xiao-Xia
Wang$^{2}$}

\affiliation{$^{1}$School of Physics, Shandong University, Jinan, Shandong 250100, China\\
$^{2}$ Institute of High Energy Physics, P.O. Box 918(4), Beijing 100049, China\\
$^{3}$ Department of Physics and Institute of Theoretical Physics,
Nanjing Normal University, Nanjing 210097, China}

\begin{abstract}
Under two different scenarios for the light scalar mesons, we
investigate the transition form factors of $B(B_s)$ mesons decay
into a scalar meson in the perturbative QCD approach. In the large
recoiling region, the form factors are dominated by the
short-distance dynamics and can be calculated using perturbation
theory. We adopt the dipole parametrization to recast the $q^2$
dependence of the form factors. Since the decay constants defined by
the scalar current are large, our predictions on the $B\to S$ form
factors are much larger than the $B\to P$ transitions, especially in
the second scenario. Contributions from various light-cone
distribution amplitudes (LCDAs) are elaborated and we find that the
twist-3 LCDAs provide more than a half contributions to the form
factors. The two terms of the twist-2 LCDAs give destructive
contributions in the first scenario while they give constructive
contributions in the second scenario. With the form factors, we also
predict the decay width and branching ratios of the semileptonic
$B\to Sl\bar\nu$ and $B\to Sl^+l^-$ decays. The branching ratios of
$B\to Sl\bar\nu$ channels are found to have the order of $10^{-4}$
while those of $B\to Sl^+l^-$ have the order of $10^{-7}$. These
predictions can be tested by the future experiments.
\end{abstract}

 \maketitle

%===============================================================================
 \section{introduction}
%===============================================================================

Although a number of scalar states have been discovered since long
time ago, the underlying structure of scalar mesons has not been
well established(for a review, see \cite{pdg}-\cite{scalars2}). In
order to uncover the inner structures, many different descriptions
have been proposed such as $\bar qq$, $\bar q\bar qqq$, meson-meson
bound states or even supplemented with a scalar glueball. It is very
likely that they are not made of one simple component but are the
superpositions of these contents. The different scenarios tend to
give very different predictions on the production and decay of the
scalar mesons which   are helpful to determine the dominant
component. Although intensive study has been given to the decay
property of the scalar mesons, the production of these mesons can
provide a different unique insight to the mysterious structure of
these mesons, especially their production in $B$ decays.

In $B$ meson decays, the energy release is much large and many
channels involving a scalar meson in the final state are open. Since
the first observation of the scalar meson $f_0(980)$ in three-body
$B$ meson decays $B^- \to K^-f_0(980)\to K^-(\pi^+\pi^-)$
\cite{Bellef0}, the two collaborations, BaBar and Belle, have
reported many studies on decays involving a scalar meson in the
final state: the branching ratios and/or direct CP asymmetries are
measured or set an upper limit~\cite{Barberio:2008fa}. Since much
more interesting channels are still not observed at present, it is
just the beginning of scalar meson study in $B$ factories.
Meanwhile, it is also necessary to provide more theoretical studies
which are useful for future experiments.

Theoretically, the studies on hadronic B decays are usually polluted
by the nonperturbative QCD effect and predictions on the observables
always suffer large uncertainties.  Since there is only one hadron
in the final state in semileptonic $B\to S$ decays, they receive
less theoretical uncertainties. In these channels, the most
challenging part in the calculation is the matrix element of the
$B_{(s)}$ to scalar meson transition.
%
%%Many approaches or models are applied to deal with this problem,
%%such as simple quark model\cite{quarkmodel}, light front
%%approach\cite{lightfront}, QCD sum rules\cite{sumrule}, light-cone
%%QCD sum rules\cite{lcsumrule}. In these studies, great discrepancies
%%are found and more studies are required.
%
%
In the region of small recoil, where $q^2$ is large, the form
factors are dominated by the soft dynamics, which is out of control
of perturbative QCD. However, in the large-recoil region where
$q^2\to 0$, roughly 5 GeV of energy is released. About half of this
energy is taken by the light scalar meson, which suggests that large
momentum is transferred in this process and the interaction is
mainly dominated by the short-distance dynamics. Therefore the
perturbative QCD approach (PQCD) \cite{pQCD} is expected to be
applicable to B to scalar meson transitions in the large-recoil
region. With the results obtained in the restricted region, one can
extrapolate these form factors  to the whole kinematic region by
adopting some parametrization form for the form factors.

This paper is organized as following: The distribution amplitudes
and decay constants of the mesons are given in Section
\ref{section:CandLCDAs}. In Section \ref{section:FFandSD} we listed
the formulae about the form factors and semileptonic decays. Section
\ref{section:results} are the discussion of the numerical results.
The Appendix \ref{Appendix:pQCDfunctions} lists out the useful
functions for PQCD approach.

 %===============================================================================
 \section{conventions and Inputs}
 \label{section:CandLCDAs}
 %===============================================================================
%
%  %-------------------------------
%  \subsection{Conventions}
%  %-------------------------------

We will work in the rest frame of the B meson and use the light-cone
coordinates. In the heavy quark limit the mass difference of b quark
and B meson is negligible: $m_b\simeq m_B$. The masses of scalar
mesons are very small compared with the b quark mass, we keep them
up to the first order. Since  the scalar meson in the final state
moves very fast in the large-recoil region, we define the momentum
of the scalar meson on the plus direction in the light-cone
coordinates. The momentum of B meson and scalar mesons can be
denoted  as
 \begin{eqnarray}
 P_B=\frac{m_B}{\sqrt{2}}(1,1,0_{\perp})\;,\;P_S=\frac{m_B}{\sqrt{2}}(\eta,0,0_{\perp})\;.\label{eq:momentum}
 \end{eqnarray}
Then for momentum $q=P_B-P_S$, there exists $\eta=1-q^2/m_B^2$. The
momentum of the light antiquark in B meson and the quark in scalar
mesons are denoted as $k_1$ and $k_2$ respectively(see
Fig.\ref{fig:form factor}):
 \begin{eqnarray}
 k_1=(0,\frac{m_B}{\sqrt{2}}x_1,\textbf{k}_{1\perp})\;,\;k_2=(\frac{m_B}{\sqrt{2}}
 x_2\eta,0,\textbf{k}_{2\perp})\;.\label{eq:fmomentum}
 \end{eqnarray}

%
% %---------------------------------------
% \subsection{Wave Function of $B_{(s)}$ Mesons}
% %---------------------------------------
In the course of the PQCD calculations,   the light-cone wave
functions of the mesons are required. The B meson is  a heavy-light
system, and its light cone matrix element can be decomposed as
\cite{Bwavefunctions}
 \begin{eqnarray}
 &&\int_0^1\frac{d^4z}{(2\pi)^4}e^{ik_1\cdot z}\langle 0|b_{\beta}(0)\bar
 q_{\alpha}(z)|\bar B_{(s)}(P_{B_{(s)}})\rangle\nonumber\\
 &=&\frac{i}{\sqrt{2N_c}}\left\{(\not\! P_{B_{(s)}}+m_{B_{(s)}})\gamma_5\left[\phi_{B_{(s)}}(k_1)+\frac{\not\!
 n-\not\!
 v}{\sqrt{2}}\bar\phi_{B_{(s)}}(k_1)\right]\right\}_{\beta\alpha}\;,\label{Bwav:decompose}
 \end{eqnarray}
where $n=(1,0,\textbf{0}_T)$ and $v=(0,1,\textbf{0}_T)$ are
light-like unit vectors. There are two Lorentz structures in B meson
light-cone distribution amplitudes, and they obey the normalization
conditions:
 \begin{equation}
 \int\frac{d^4 k_1}{(2\pi)^4}\phi_{B_{(s)}}({
 k_1})=\frac{f_{B_{(s)}}}{2\sqrt{2N_c}}\;,\; \int \frac{d^4
 k_1}{(2\pi)^4}\bar{\phi}_{B_{(s)}}({ k_1})=0,\label{Bwave:normalization}
 \end{equation}
with $f_{B_{(s)}}$ as the decay constant of $B_{(s)}$ meson. In
principle, both the $\phi_{B_{(s)}}(k_1)$ and
$\bar\phi_{B_{(s)}}(k_1)$ contribute in B meson transitions.
However, the contribution of $\bar \phi_{B_{(s)}}(k_1)$ is usually
neglected, because its contribution is numerically small
\cite{Lu:Btolight}.
 So we will only keep the term with $\phi_{B_{(s)}}(k_1)$ in equation
 (\ref{Bwav:decompose}). In the momentum space the light cone matrix of B
meson can be expressed as:
 \begin{equation}
 \Phi_{B_{(s)}}=\frac{i}{\sqrt{6}}(\not \! P_{B_{(s)}} +m_{B_{(s)}})\gamma_5\phi_{B_{(s)}} (k_1). \label{Bwave:3variable}
 \end{equation}
Usually the hard part is independent of $k^+$ or/and $k^-$, so we
integrate one of them out from
$\phi_{B_{(s)}}(k^+,k^-,\textbf{k}_{\perp})$. With
 $b$ as the conjugate space coordinate of $\textbf{k}_{\perp}$, we can express
 $\phi_{B_{(s)}}(x,\textbf{k}_{\perp})$ in b-space by
 \begin{equation}
 \Phi_{{B_{(s)}},\alpha\beta}(x,b) = \frac{i}{\sqrt{2N_c}}
 \left[\not\! P_{B_{(s)}} \gamma_5 + m_{B_{(s)}}\gamma_{5} \right]_{\alpha\beta}
 \phi_{B_{(s)}}(x,b),\label{Bwave:bspace}
 \end{equation}
where $x$ is the momentum fraction of the light quark in B meson. In
this paper, we use the following expression for
$\phi_{B_{(s)}}(x,b)$:
 \begin{equation}
 \phi_{B_{(s)}}(x,b)=N_{B_{(s)}}x^2(1-x)^2\mbox{exp}\left[-\frac{m_{B_{(s)}}^2 x^2}{2\omega_b^2}-\frac{1}{2}(\omega_b
 b)^2\right], \label{Bwave:da}
 \end{equation}
with $N_{B_{(s)}}$ the normalization factor, which is determined by
equation (\ref{Bwave:normalization}). In recent years, a lot of
studies for $B^{\pm}$ and $B_d^0$ decays have been performed by PQCD
approach \cite{pQCD}. With the rich experimental data, the
$\omega_b$ in (\ref{Bwave:da}) is fixed as $0.40\mbox{GeV}$. In our
calculation, we adopt $\omega_b=(0.40\pm0.05)\mbox{GeV}$ and
$f_B=(0.19\pm0.025\rm{GeV})$ for B mesons. For $B_s$ meson, taking
the SU(3) breaking effects into consideration, we adopt
$\omega_{b_s}=(0.50\pm 0.05)\mbox{GeV}$ and
$f_{B_s}=0.23\pm0.03\rm{GeV}$\cite{Lu:Bs}.
%
%%-------------------------------------
%\subsection{Scalar Mesons}
%%-------------------------------------
%\subsubsection{Two Scenarios}

In the spectroscopy study, many scalar states have been discovered.
Among them, the scalar mesons below 1 GeV, including
$f_0(600)(\sigma)$, $f_0(980)$, $K^*_0(800)(\kappa)$ and $a_0(980)$,
are usually viewed to form an SU(3) nonet; while scalar mesons
around 1.5 GeV, including $f_0(1370)$, $f_0(1500)$/$f_0(1700)$,
$K^*_0(1430)$ and $a_0(1450)$, form another nonet. There are two
different scenarios to describe these mesons in the quark model. The
first one(called scenario 1 in this paper) is the naive 2-quark
model: the nonet mesons below 1 GeV are treated as the lowest lying
states, and the ones near 1.5 GeV are the first excited state.  In
this scenario, the flavor wave functions of the light scalar mesons
are
 \begin{eqnarray}
 \sigma=\frac{1}{\sqrt{2}}(u\bar u+d\bar d),\;\;f_0=s\bar s,\nonumber\\
 a_0^+=u\bar d,\;a_0^0=\frac{1}{\sqrt{2}}(u\bar u+d\bar d)\;,a_0^-=d\bar
 u, \nonumber\\
 \kappa^+=u\bar s,\;\kappa^0=d\bar s,\;\bar\kappa^0=s\bar
 d,\;\kappa^-=s\bar u.
 \end{eqnarray}
Here it's supposed that the $\sigma$ and $f_0(980)$ has the ideal
mixing. However, the data of $J/\psi$ decays doesn't favor
$f_0(980)$ as a pure $s\bar s$ state\cite{HaiYang-Chen:scalar}, and
it seems  that $\sigma$ and $f_0(980)$ have a mixing like
 \begin{eqnarray}
 &&|f_0(980)\rangle=|s\bar s\rangle \cos\theta +|n\bar n\rangle
 \sin\theta,\nonumber\\
 &&|\sigma\rangle=-|s\bar s\rangle \sin\theta+|n\bar n\rangle
 \cos\theta,
 \end{eqnarray}
with $|n\bar n\rangle=\frac{1}{\sqrt{2}}(u\bar u+d\bar d)$ and
$\theta$ as the mixing angle.  The above description has encountered
several severe difficulties. For example, if the $\bar qq$ states
have the quantum numbers $J^{PC}=0^{++}$, the corresponding masses
are expected larger than that of the vector mesons. Studies on the
mixing angle of $\sigma$ and $f_0(980)$\cite{mixingangle} show that
$\theta$ tends to be not a unique value, which indicates that
$\sigma$ and $f_0(980)$ may not be purely $q\bar q$ states. Based on
these facts, the second scenario is proposed, where the nonet mesons
near 1.5GeV are viewed as the lowest lying states, while the mesons
below 1 GeV may be viewed as four-quark bound states. Because of the
difficulty when dealing with four-quark states, we only do the
calculation about the heavier nonet in this scenario.
%
%%-------------------------------------------------------------
% \subsubsection{Distribution Amplitudes of Scalar Mesons}
%%-------------------------------------------------------------

The decay constants of scalar mesons are defined
by\cite{HaiYang-Chen:scalar}
 \begin{eqnarray}
 \langle S(p)|\bar
 q_2\gamma_{\mu}q_1|0\rangle=f_Sp_{\mu}\;,\; \langle S|\bar q_2q_1|0\rangle=m_S\bar f_S.
 \end{eqnarray}
Because of the charge conjugate invariance, neutral scalar mesons
cannot be produced by the vector current and thus
 \begin{equation}
 f_{\sigma}=f_{f_0}=f_{a_0^0}=0.
 \end{equation}
For other scalar mesons, the vector decay constant $f_S$ and scalar
decay constant $\bar f_S$(listed in Table \ref{tab:Gmoments} and
\ref{tab:Gmoments2}) is related by equations of motion
$\mu_sf_S=\bar f_S$, with $\mu_s=\frac{m_S}{m_2(\mu)-m_1(\mu)}$.
$m_S$ is the mass of the scalar meson, and $m_1$, $m_2$ are the
running current quark masses. Inputs of the scalar mesons in our
calculation, include the decay constants, running quark masses in
this paragraph and the Gegenbauer moments in the following, quote
from \cite{HaiYang-Chen:scalar}.

The definition of twist-2 light-cone distribution amplitude(LCDA)
$\Phi_S(x)$ and twist-3 LCDAs $\Phi_S^s(x)$ and $\Phi_S^{\sigma}$
for the scalar mesons can be combined into a single matrix
element\cite{HaiYang-Chen:scalar}:
 \begin{eqnarray}
 \langle S(P_S)|q(0)_j\bar q(z)_l |0\rangle
 &=&\frac{-1}{\sqrt{2N_c}}\int^1_0dxe^{ixp\cdot
 z}\{\not P_S\Phi_S(x)
 +m_S\Phi^s_S(x)+m_S\sigma_{\mu\nu}P_S^{\mu}z^{\nu}\frac{\Phi^{\sigma}_S(x)}{6}\}_{jl}\nonumber\\
 &=&\frac{-1}{\sqrt{2N_c}}\int^1_0dxe^{ixp\cdot
 z}\{\not P_S\Phi_S(x)
 +m_S\Phi^s_S(x)+m_S(\not{n}\not{v}-1)\Phi^T_S(x)\}_{jl},\label{LCDA}
 \end{eqnarray}
with the normalization conditions
 \begin{eqnarray}
 \int_0^1 dx\phi_S(x)&=&\frac{f_S}{2\sqrt{2N_c}},\nonumber\\
 \int_0^1 dx\phi_S^s(x)&=&\int_0^1
 dx\phi_S^{\sigma}(x)=\frac{\bar f_S}{2\sqrt{2N_c}}.
 \end{eqnarray}
The LCDAs can be expanded in Gegenbauer polynomials as the following
form:
 \begin{eqnarray}
 \phi_S(x)&=&\frac{f_S}{2\sqrt{2N_c}}6x(1-x)\bigg[1+\mu_s\sum_{m=1}^{\infty}B_m(\mu)C_m^{{3/2}}(2x-1)\bigg],\nonumber\\
 &=&\frac{\bar
 f_S}{2\sqrt{2N_c}}6x(1-x)\bigg[\frac{1}{\mu_s}+\sum_{m=1}^{\infty}B_m(\mu)C_m^{{3/2}}(2x-1)\bigg],\\
 \phi_S^s(x)&=&\frac{\bar f_S}{2\sqrt{2N_c}}[1+\sum_{m=1}^{\infty}a_m(\mu)C_m^{1/2}(2x-1)],\\
 \phi_S^T(x)&=&\frac{d}{dx}\frac{\phi_S^{\sigma}(x)}{6}=\frac{\bar f_S}{2\sqrt{2N_c}}\frac{d}{dx}\bigg\{
 x(1-x)[1+\sum_{m=1}^{\infty}b_m(\mu)C_m^{3/2}(2x-1)]\bigg\}.
 \end{eqnarray}
where $B_m(\mu)$, $a_m(\mu)$ and $b_m(\mu)$ are the Gegenbauer
moments and $C_m^{(3/2)}$ and $C_m^{1/2}$ are the Gegenbauer
polynomials. The values of $B_m(\mu)$ are listed in Table
\ref{tab:Gmoments} and \ref{tab:Gmoments2}. And the values of
$b_m{\mu}$ and $a_m(\mu)$ in scenario 2 are worked out in
\cite{Lu:2006fr}, which is listed in Table \ref{tab:Gtwist3}.
However, in the calculation in scenario 1, the asymptotic form of
twist-3 LCDAs is used.

%******************************************************************************************************
%******************************************************************************************************
% \begin{table}
% \caption{Decay constants($\bar f_S$) at scale $\mu=1\rm{GeV}$ in scenario 1(Unit:$\rm{MeV}$).}
% \label{tab:decayCs}
% \begin{center}
% \begin{tabular}{c c c c c c}
% \hline\hline
% \hline
% \ \ \     $a_0(980)$       &$a_0(1450)$       &$f_0(980)$      &$f_0(1500)$       &$\kappa(800)$      &$K^*_0(1430)$   \\
% \ \ \     $365\pm 20$      &$-280\pm 30$      &$370\pm 20$     &$-255\pm 30$      &$340\pm 20$        &$-300\pm 30$   \\
% \hline\hline
% \end{tabular}
% \end{center}
% \end{table}

%******************************************************************************************************
%******************************************************************************************************
% \begin{table}
% \caption{Decay constants($\bar f_S$) at scale $\mu=1\rm{GeV}$ in scenario 2(Unit:$\rm{MeV}$).}
% \label{tab:decayCs2}
% \begin{center}
% \begin{tabular}{c c c}
% \hline\hline
% \hline
% \ \ \     $a_0(1450)$     &$f_0(1500)$      &$K^*_0(1430)$   \\
% \ \ \     $460\pm 50$     &$490\pm 50$      &$445\pm 50$   \\
% \hline\hline
% \end{tabular}
% \end{center}
% \end{table}

%******************************************************************************************************
%******************************************************************************************************
 \begin{table}
 \caption{Decay constants $\bar f_S$ (in unit of MeV)  and Gegenbauer moments at scale $\mu=1\rm{GeV}$ in scenario 1.}
 \label{tab:Gmoments}
 \begin{center}
 \begin{tabular}{c|c c c}
 \hline\hline
 \ \ \         & $\bar f_S$ &$B_1$           &$B_3$        \\
 \hline
 \ \ \     $a_0(980)$       & $365\pm20$ &$-0.93\pm 0.10$       &$0.14\pm 0.08$               \\
 \ \ \     $a_0(1450)$      & $-280\pm30$ &$0.89\pm 0.20$       &$-1.38\pm 0.18$               \\
 \ \ \     $f_0(980)$       &  $370\pm20$ &$-0.78\pm 0.08$       &$0.02\pm 0.07$               \\
 \ \ \     $f_0(1500)$      & $-255\pm30$ &$0.80\pm 0.40$       &$-1.32\pm0.14$               \\
 \ \ \     $\kappa(800)$    & $340\pm20$ &$-0.92\pm 0.11$       &$0.15\pm 0.09$               \\
 \ \ \     $K^*_0(1430)$    & $-300\pm30$ &$0.58\pm 0.07$       &$-1.20\pm 0.08$               \\
 \hline\hline
 \end{tabular}
 \end{center}
 \end{table}

%******************************************************************************************************
%******************************************************************************************************
 \begin{table}
 \caption{Decay constants $\bar f_S$ (in unit of MeV) and Gegenbauer moments at scale $\mu=1\rm{GeV}$ in scenario 2.}
 \label{tab:Gmoments2}
 \begin{center}
 \begin{tabular}{c|c c c}
 \hline\hline
 \ \ \         & $\bar f_S$ &$B_1$           &$B_3$        \\
 \hline
 \ \ \     $a_0(1450)$     &$460\pm 50$  &$-0.58\pm 0.12$       &$-0.49\pm 0.15$               \\
 \ \ \     $f_0(1500)$     & $490\pm50$  &$-0.48\pm 0.11$       &$-0.37\pm 0.20$               \\
 \ \ \     $K^*_0(1430)$   & $445\pm50$  &$-0.57\pm 0.13$       &$-0.42\pm 0.22$               \\
 \hline\hline
 \end{tabular}
 \end{center}
 \end{table}
%******************************************************************************************************
%******************************************************************************************************

%******************************************************************************************************
%******************************************************************************************************
 \begin{table}
 \caption{Gegenbauer moments for the twist-3 LCDAs of scalar mesons at the scale $\mu=1\rm{GeV}$ in scenario
 2\cite{Lu:2006fr}.}
 \label{tab:Gtwist3}
 \begin{center}
 \begin{tabular}{c c c c c c c}
 \hline\hline
 \ \ \     state    & $a_1(\times 10^{-2})$     &$a_2$     &$a_4$      &$b_1(\times 10^{-2})$     &$b_2$     &$b_4$\\
 \hline
 \ \ \     $a_0(1450)$     &$0$       &$-0.33\sim -0.18$       &$-0.11\sim 0.39$     &$0$       &$0\sim 0.058$       &$0.070\sim 0.20$\\
 \ \ \     $K^*_0(1430)$   &$1.8\sim 4.2$       &$-0.33\sim -0.025$       &$-$     &$3.7\sim 5.5$       &$0\sim 0.15$       &$-$\\
 \ \ \     $f_0(1500)$     &$0$       &$-0.33\sim 0.18$       &$0.28\sim 0.79$     &$0$       &$-0.15\sim -0.088$       &$0.044\sim 0.16$\\
 \hline\hline
 \end{tabular}
 \end{center}
 \end{table}
%******************************************************************************************************
%******************************************************************************************************

%===========================================================================
 \section{$B\to S$ Form Factors and Semileptonic Decays in the PQCD approach}
 \label{section:FFandSD}
%===========================================================================
%------------------------------------
\subsection{A Brief Review of pQCD Approach}
%------------------------------------
The basic idea of pQCD approach is including the intrinsic
transverse momenta of valence quarks in the calculation of the
hadronic matrix elements. The transition matrix element(see Fig.
\ref{fig:form factor}) of B meson to a scalar meson($q_1\bar q_2$
component is supposed) can be expressed as the convolution of the
wave functions $\Phi_B$, $\Phi_S$ and the hard scattering kernel
$T_H$, integrated over the longitudinal and transverse momenta of
the valence quarks:
 \begin{eqnarray}
 {\cal M}\propto \int_0^1dx_1dx_2\int_{-\infty}^{\infty} \frac{d^2\vec{k}_{1\perp}}{(2\pi)^2}
 \frac{d^2\vec{k}_{2\perp}}{(2\pi)^2}\Phi_B(x_1,\vec{k}_{1\perp},p_B,t)
 T_H(x_1,x_2,\vec{k}_{1\perp},\vec{k}_{2\perp},t)\Phi_S(x_2,\vec{k}_{2\perp},p_1,t).
 \label{eq:pQCDfactorization}
 \end{eqnarray}
It's convenient to calculate the transition amplitude in coordinate
space. Through the Fourier transformation, the above equation
becomes
 \begin{eqnarray}
 {\cal M}\propto \int_0^1dx_1dx_2 \int_{-\infty}^{\infty}d^2\vec{b}_1d^2\vec{b}_2\Phi_B(x_1,\vec{b}_1,p_B,t)
 T_H(x_1,x_2,\vec{b}_1,\vec{b}_2,t)\Phi_S(x_2,\vec{b}_2,p_1,t).
 \end{eqnarray}
In principle, loop corrections to scattering kernel $T_H$ can be
taken into consideration, which usually bring two types of infrared
divergences in individual diagrams: soft and collinear. Soft
divergence is generated when all the components of a loop momentum
$l$ go to zero:
 \begin{equation}
 l^{\mu}=(l^+,l^-,\vec{l}_T)=(\Lambda,\Lambda,\vec{\Lambda}),
 \end{equation}
with $l^{\mu}$ expressed in the light-cone coordinate. The collinear
divergence arise from the region where the gluon momentum is
parallel to the massless quark momentum:
 \begin{equation}
 l^{\mu}=(l^+,l^-,\vec{l}_T)=(m_B,\Lambda^2/m_B,\vec{\Lambda}).
 \end{equation}
In both cases, the loop integration correspond to $\int d^4l/l^4
\sim \log\Lambda$, thus logarithmic divergences are generated. In
perturbation theory, it has been shown order by order that these
divergences can be separated from the hard kernel and obsorbed into
meson wave functions using eikonal approximation\cite{eikonal}. When
the soft and collinear momenta overlap, one also encounter double
logarithm divergences, which can be resummed into the Sudakov factor
and its expression is given in Appendix
\ref{Appendix:pQCDfunctions}.

The loop corrections to the weak decay vertex will generate another
type of double logarithm. For example, the amplitude of the left
diagram of Fig. \ref{fig:form factor} is proportional to
$1/(x_2^2x_1)$. When $x_2\to 0$, additional collinear divergences
are associated with the internal quark. The integration of the
amplitude will produce double logarithm $\alpha_s\ln^2x_2$, and the
resummation of this type of double logarithm gives rise to Sudakov
factor $S_t(x_2)$\cite{jetfunction1}, which is usually called jet
function. The similar jet function $S_t(x_1)$ is generated after the
resummation of the same type of double logarithm of the right
diagram in Fig. \ref{fig:form factor}. The jet function decreases
faster than any power of $x$ as $x\to 0$, thus it kills the endpoint
singularity effectively. The jet function has been parametrized in a
form which is independent of the decay channels, twists and
flavors\cite{jetfunction2}.

With the Sudakov factors included, the factorization formula of the
form factor matrix element in pQCD approach is given by
 \begin{eqnarray}
 {\cal M}&\propto &\int_0^1dx_1dx_2 \int_{-\infty}^{\infty}d^2\vec{b}_1d^2\vec{b}_2\Phi_B(x_1,\vec{b}_1,p_B,t)
 T_H(x_1,x_2,\vec{b}_1,\vec{b}_2,t)\nonumber\\
 &&\times\Phi_S(x_2,\vec{b}_2,p_1,t)S_t(x_i)\rm{exp}[-S_B(t)-S_2(t)].
 \end{eqnarray}

%------------------------------------
\subsection{Form Factors in the PQCD approach}
%------------------------------------
 \begin{figure}
 \begin{center}
 \includegraphics[scale=1]{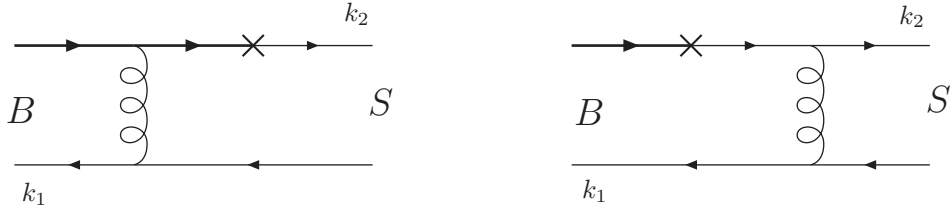}
 \caption{Contributions to the form factors in the PQCD approach, where the cross
 denotes the weak vertex.}
 \label{fig:form factor}
 \end{center}
 \end{figure}
 The form factors for $B_{(s)}\to S$ transition are defined by
  \begin{eqnarray}
   \kappa_S \langle S(P_S) |\bar q  \gamma_\mu\gamma_5  b|\overline  {B_{(s)}}(P_{B_{(s)}})\rangle
    &=&-i\left\{\left
    [(P_{B_{(s)}} +P_S)_\mu - \frac{m_{B_{(s)}}^2-m_S^2}{q^2}q_\mu \right ]
    F_1 (q^2)
    +\frac{m_{B_{(s)}}^2-m_S^2}{q^2}q_\mu F_0 (q^2)\right\} ,\\
%-------------------------------------------------------------------------------
   \kappa_S\langle S(P_S) |\bar q  \sigma_{\mu\nu} b|\overline  {B_{(s)}}(P_{B_{(s)}})\rangle
   &=&-i\epsilon_{\mu\nu\alpha\beta}p_1^{\alpha}q^{\beta}
  \frac{2 F_T (q^2)}{
   m_{B_{(s)}}+m_S},\\
%-------------------------------------------------------------------------------
   \kappa_S\langle S(P_S) |\bar q  \sigma_{\mu\nu}\gamma_5  b|\overline  {B_{(s)}}(P_{B_{(s)}})\rangle &=&\left
   [q_\mu {P_S}_\nu - {P_S}_\mu  q_\nu  \right ]
  \frac{2 F_T (q^2)}{
   m_{B_{(s)}}+m_S},
   \end{eqnarray}
with $q=P_{B_{(s)}}-P_S$. $\kappa_S$ is the flavor factor for the
transition: $\sqrt 2$ for the component of $\bar uu$ in the
$\frac{\bar uu\pm \bar dd}{\sqrt 2}$ state, $\pm \sqrt 2$ for the
component of $\bar dd$ in the $\frac{\bar uu\pm \bar dd}{\sqrt 2}$
state, 1 for the other states. In the large-recoil region, a hard
gluon is required to kick the soft spectator  antiquark to a
fast-moving antiquark. Therefore, in this kinematics region, the
form factors can be calculated perturbatively. The lowest order
diagrams for the $B_{(s)}\to S$ transition are shown in
Fig.\ref{fig:form factor}. Carrying out the calculation under pQCD
approach, we obtain the analytic formulae of the form factors nearby
the $q^2=0$:
 \begin{eqnarray}
 F_0(\eta)&=&8 \pi C_F m_B^2 \int_0^1
        dx_1dx_2\int_0^{\infty}b_1db_1b_2db_2 \phi_B(x_1,b_1)\nonumber\\
        &&\times\left\{\big[\eta(x_2\eta-\eta-1)\phi_S(x_2)-r_2\eta(2x_2-1)\phi_S^T(x_2)
        +r_2(2x_2\eta-3\eta+2)\phi_S^s(x_2)\big]\right.\nonumber\\
        &&\left.\times h_e(x_1,(1-x_2)\eta,b_1,b_2)\alpha_s(t_e^1)\mbox{exp}[-S_{ab}(t_e^1)]S_t(x_2)\right.\nonumber\\
        &&\left.+2r_2\eta\phi_S^s(x_2)h_e(1-x_2,x_1\eta,b_2,b_1)\alpha_s(t_e^2)\mbox{exp}[-S_{ab}(t_e^2)]S_t(x_1)\right\},\label{eq:f0}\\
 %------------------------------------------------------------------------------------------------------------------
 F_1(\eta)&=&8 \pi C_F m_B^2 \int_0^1
        dx_1dx_2\int_0^{\infty}b_1db_1b_2db_2 \phi_B(x_1,b_1)\nonumber\\
        &&\times\left\{\big[(x_2\eta-\eta-1)\phi_S(x_2)+r_2(-2x_2+3-2/\eta)\phi_S^T(x_2)
        -r_2(1-2x_2)\phi_S^s(x_2)\big]\right.\nonumber\\
        &&\left.\times h_e(x_1,(1-x_2)\eta,b_1,b_2)\alpha_s(t_e^1)\mbox{exp}[-S_{ab}(t_e^1)]S_t(x_2)\right.\nonumber\\
        &&\left.+2r_2\phi_S^s(x_2)h_e(1-x_2,x_1\eta,b_2,b_1)\alpha_s(t_e^2)\mbox{exp}[-S_{ab}(t_e^2)]S_t(x_1)\right\},\\
 %---------------------------------------------------------------------------------------------------------------------
 F_T(\eta)&=&8 \pi C_F m_B^2 (1+r_2)\int_0^1 dx_1dx_2 \int_0^{\infty} b_1 db_1 b_2 db_2 \phi_B(x_1,b_1) \nonumber\\
        &&\times\left\{ \big[r_2(x_2-1)\phi_S^s(x_2)-\phi_S(x_2)+r_2(x_2-1-2/\eta)\phi_S^T(x_2)\big]\right.\nonumber\\
        &&\left.\times h_e(x_1,(1-x_2)\eta,b_1,b_2)\alpha_s(t_e^1)\mbox{exp}[-S_{ab}(t_e^1)]S_t(x_2)\right.\nonumber\\
        &&\left.+2r_2\phi_S^s(x_2)
        h_e(1-x_2,x_1\eta,b_2,b_1)\alpha_s(t_e^2)\mbox{exp}[-S_{ab}(t_e^2)]S_t(x_1)\right\}.\label{eq:fT}
 \end{eqnarray}
With these formulae we calculate the form factors nearby $q^2=0$.
Through fitting the results among the region $0<q^2<10\mbox{GeV}^2$,
we extrapolate them with the pole model parametrization
 \begin{equation}
 F_i(q^2)=\frac{F_i(0)}{1-a(q^2/m_B^2)+b(q^2/m_B^2)^2},
 \end{equation}
 with $a,b$ are the constants to be determined from the fitting procedure.

 %------------------------------------------
 \subsection{Semileptonic $B_{(s)}$ Meson decays }
 %-------------------------------------------
% \subsubsection{Effective Hamiltonian for $b\to ul\bar \nu_l$ transition}
The effective Hamiltonian for $b\to ul\bar \nu_l$ transition is
 \begin{eqnarray}
 {\cal H}_{eff}(b\to ul\bar \nu_l)=\frac{G_F}{\sqrt{2}}V_{ub}\bar
 u\gamma_{\mu}(1-\gamma_5)b \bar l\gamma^{\mu}(1-\gamma_5)\nu_l.
 \end{eqnarray}
With the Hamiltonian, the $q^2$ dependant decay width
$\frac{d\Gamma}{dq^2}$ can be expressed as
 \begin{eqnarray}
 \frac{d\Gamma}{dq^2}&=&\frac{G_F^2|V_{ub}|^2}{192 \pi^3
 m_B^3}\frac{q^2-m_l^2}{(q^2)^2}\sqrt{\frac{(q^2-m_l^2)^2}{q^2}}
 \sqrt{\frac{(m_B^2-m_S^2-q^2)^2}{4q^2}-m_S^2}\nonumber\\
 &&\bigg[(m_l^2+2q^2)(q^2-(m_B-m_S)^2)(q^2-(m_B+m_S)^2)F_1^2(q^2)+3m_l^2(m_B^2-m_S^2)^2F_0^2(q^2)\bigg],\label{eq:pwel}
 \end{eqnarray}
with $m_l$ as the mass of the lepton.

The calculation of $b\to sl^+l^-$ transition is a bit complicated,
because both the short-distance and long-distance contribution
should be taken into consideration. The weak effective Hamiltonian
is
 \begin{eqnarray}
 {\cal
 H}_{\mbox{eff}}=-\frac{G_F}{\sqrt{2}}V_{tb}V^*_{ts}\sum_{i=1}^{10}C_i(\mu)O_i(\mu),\label{eq:Hamiltonian}
 \end{eqnarray}
with the doubly CKM suppressed terms omitted. $C_i(\mu)$ are the
Wilson coefficients and the local operators $O_i(\mu)$ are given by
\cite{operators}
 \begin{eqnarray}
 O_1&=&(\bar s_{\alpha}c_{\alpha})_{V-A}(\bar
 c_{\beta}b_{\beta})_{V-A},\;\;
 O_2=(\bar
 s_{\alpha}c_{\beta})_{V-A}(\bar
 c_{\beta}b_{\alpha})_{V-A},\nonumber\\
 %-------------------------------------------------
 O_3&=&(\bar s_{\alpha}b_{\alpha})_{V-A}\sum_q(\bar
 q_{\beta}q_{\beta})_{V-A},\;\;
 O_4=(\bar s_{\alpha}b_{\beta})_{V-A}\sum_q(\bar
 q_{\beta}q_{\alpha})_{V-A},\nonumber\\
 %-------------------------------------------------
 O_5&=&(\bar s_{\alpha}b_{\alpha})_{V-A}\sum_q(\bar
 q_{\beta}q_{\beta})_{V+A},\;\;
 O_6=(\bar s_{\alpha}b_{\beta})_{V-A}\sum_q(\bar
 q_{\beta}q_{\alpha})_{V+A},\nonumber\\
 %-------------------------------------------------
 O_7&=&\frac{e m_b}{8\pi^2}\bar
 s\sigma^{\mu\nu}(1+\gamma_5)bF_{\mu\nu},\nonumber\\
 %------------------------------------------------------
 O_9&=&\frac{\alpha_{\rm{em}}}{8\pi}(\bar l\gamma_{\mu}l)(\bar
 s\gamma^{\mu}(1-\gamma_5)b),\;\;
 O_{10}=\frac{\alpha_{\rm{em}}}{8\pi}(\bar l\gamma_{\mu}\gamma_5l)(\bar
 s\gamma^{\mu}(1-\gamma_5)b),\label{eq:operators}
 \end{eqnarray}
where $(\bar q_1q_2)_{V-A}(\bar q_3 q_4)_{V-A}\equiv(\bar q_1
\gamma^{\mu}(1-\gamma_5)q_2)(\bar q_3\gamma_{\mu}(1-\gamma)q_4)$,
and $(\bar q_1q_2)_{V-A}(\bar q_3 q_4)_{V+A}\equiv(\bar q_1
\gamma^{\mu}(1-\gamma_5)q_2)(\bar q_3\gamma_{\mu}(1+\gamma)q_4)$. In
equation (\ref{eq:operators}), the term suppressed by $m_s$ in $O_7$
is neglected.

The amplitude for $b\to sl^+l^-$ transition can be decomposed as
 \begin{eqnarray}
 {\cal A}(b\to sl^+
 l^-)&=&\frac{G_F}{2\sqrt{2}}\frac{\alpha_{\rm{em}}}{\pi}V^*_{ts}V_{tb}\bigg\{
 C_9^{eff}(\mu)
 [\bar s \gamma_{\mu}(1-\gamma_5)b][\bar l\gamma^{\mu}l] + C_{10}[\bar s\gamma_{\mu}(1-\gamma_5)b]
 [\bar l\gamma^{\mu}\gamma_5l]\nonumber\\
 &&- 2m_bC_7^{eff}(\mu)\big[\bar s i\sigma_{\mu\nu}\frac{q^{\nu}}{q^2}
 (1+\gamma_5)b[\bar l\gamma^{\mu}l\big] \bigg\},
 \end{eqnarray}
where $\hat{s}=q^2/m_B^2$ and $\hat{m}_b=m_b/m_B$, with $m_b$ as the
b quark mass in the $\overline{\mbox{MS}}$ scheme. The long-distance
and short-distance contributions are absorbed into the
$C_7^{eff}(\mu)$ and $C_9^{eff}(\mu)$, with
 \begin{eqnarray}
 C_7^{\rm{eff}}(\mu)&=&C_7(\mu)+C^{\prime}_{b\to
 s\gamma}(\mu),\nonumber\\
 C_9^{\rm{eff}}(\mu)&=&C_9(\mu)+Y_{\rm{pert}}(\hat{s})+Y_{\rm{LD}}(\hat{s}).
 \end{eqnarray}
$Y_{\rm{pert}}$ represents the perturbative contributions, and
$Y_{\rm{LD}}$ is the long-distance part. The $Y_{\rm{pert}}$ is
given by\cite{ypert}
 \begin{eqnarray}
 Y_{\rm{pert}}(\hat{s})&=&
 h(\hat{m_c},\hat{s})C_0-\frac{1}{2}h(1,\hat{s})(4C_3+4
 C_4+3C_5+C_6)\nonumber\\
 &&-\frac{1}{2}h(0,\hat{s})(C_3+3
 C_4) + \frac{2}{9}(3C_3 + C_4 +3C_5+ C_6),\label{eq:ypert}
 \end{eqnarray}
with $C_0=C_1+3C_2+3C_3+C_4+3C_5 +C_6$. The Wilson coefficients,
listed in table \ref{tab:wilsons}, are given in the leading
logarithmic accuracy. The long-distance part $Y_{\rm{LD}}$,
involving the contributions of $B_{(s)}\to SV(c\bar c)$ resonances
where $V(c\bar c)$ are the charmonium states, is neglected in this
paper because of the lack of the experimental data. The corrections
of the nonfactorizable effects of the charm quark loop to the $b\to
s\gamma$ transition at $q^2=0$ are also neglected. And the
absorptive part of $b\to s\gamma$ with neglecting the small
contribution from $V_{tb}V^*_{ts}$ is represented by the
$C^{\prime}_{b \to s\gamma}$ part in $C_7^{eff}$, which is given
by(for a complete expression of $C_7^{eff}(\mu)$, see
\cite{c7nonpert})
 \begin{eqnarray}
 C^{\prime}_{b \to
 s\gamma}(\mu)=i\alpha_s[\frac{2}{9}\eta^{14/23}(G_I(x_t)-0.1687)-0.03C_2(\mu)],
 \end{eqnarray}
with
$G_I(x_t)=\frac{x_t(x_t^2-5x_t-2)}{8(x_t-1)^3}+\frac{3x_t^2ln^2x_t}{4(x_t-1)^4}$,
 $\eta=\alpha_s(m_W)/\alpha_s(\mu)$ and $x_t=m_t^2/m_W^2$.

The $q^2$ dependant width of $B\to Sl^+l^-$ is given by
% \begin{eqnarray}
% \frac{d\Gamma}{dq^2}&=&\frac{\alpha^2_{\mbox{em}} |V_{tb}|^2|V^*_{ts}|^2 G_F^2\sqrt{q^2-4m_l^2}
% \sqrt{\frac{(m_B^2-m_S^2-q^2)^2}{4q^2}-m_S^2}}{768 \pi^5 q^2
% m_B^3}\nonumber\\
% &&\left\{\bigg[\bigg((6 F_0^2(q^2)-4F_1^2(q^2))m_B^4+4(2q^2F_1^2(q^2)\right.\nonumber\\
% &&\left.+(2F_1^2(q^2)-3F_0^2(q^2))m_S^2)m_B^2+(6F_0^2(q^2)-4F_1^2(q^2))m_S^4-4F_1^2(q^2)q^4
% +8F_1^2(q^2)q^2m_S^2\bigg)m_l^2\right.\nonumber\\
% &&\left.+F_1^2(q^2)q^2(m_B^4-2(m_S^2+q^2)m_B^2+(q^2-m_S^2)^2)\bigg]C_{10}^2\right.\nonumber\\
% &&\left.+\frac{1}{(m_B+m_S)^2}\bigg[\bigg(4F_T^2(q^2)(C_7^{\mbox{eff}})^2m_b^2+4F_1(q^2)F_T(q^2)C_7^{\mbox{eff}}Re[C_9^{\mbox{eff}}]
% (m_B+m_S)m_b\right.\nonumber\\
% &&\left.+F_1^2(q^2)|C_9^{\mbox{eff}}|^2(m_B+m_S)^2\bigg)(2m_l^2+q^2)\big(m_B^4-2(m_S^2+q^2)m_B^2+(q^2-m_S^2)^2\big)\bigg]\right\}.
% \end{eqnarray}

 \begin{eqnarray}
 \frac{d\Gamma}{dq^2}&=&\frac{G_F^2 \alpha^2_{em} |V_{tb}|^2|V^*_{ts}|^2 \sqrt{\lambda}}{1024 m_B^3\pi^5}
 \sqrt{\frac{q^2-4m_l^2}{q^2}}\nonumber\\
 &&\times\left[\frac{4}{3}\lambda\bigg|\frac{C_9^{\rm{eff}}}{2}F_1(q^2)+\frac{C_{10}}{2}F_1(q^2)\sqrt{\frac{q^2-4m_l^2}{q^2}}
 +C_7^{\rm{eff}}\frac{m_b F_T(q^2)}{m_B+m_S}\bigg|^2\right.\nonumber\\
 &&\left.+\frac{4}{3}\lambda\bigg| \frac{C_9^{\rm{eff}}}{2}F_1(q^2)-\frac{C_{10}}{2}F_1(q^2)\sqrt{\frac{q^2-4m_l^2}{q^2}}
 +C_7^{\rm{eff}}\frac{m_b F_T(q^2)}{m_B+m_S}\bigg|^2\right.\nonumber\\
 &&+\left.\frac{4\lambda}{3q^2}\bigg|C_9^{\rm{eff}} m_l F_1(q^2)+C_7^{\rm{eff}}
 \frac{2 m_l^2 m_b F_T(q^2)}{m_B+m_S}\bigg|^2+4\bigg|m_l
 C_{10}(m_B^2-m_S^2)F_0(q^2)\bigg|^2\right],\label{eq:pwee}
 \end{eqnarray}
with $\lambda=(m_B^2-q^2-m_S^2)^2-4m_S^2q^2$.

%%%%%%%%%%%%%%%%%%%%%%%%%%%%%%%%%%%%%%%%%%%%%%%%%%%%%%%%%%%%%%%%%%%%%%%%%%%%%%%%%%%%%%%%%%%%%%%%%%%%%%%%%%%%%%%%
%%%%%%%%%%%%%%%%%%%%%%%%%% values of wilson coefficients %%%%%%%%%%%%%%%%%%%%%%%%%%%%%%%%%%%%%%%%%%%%%%%%%%%%%%%
%%%%%%%%%%%%%%%%%%%%%%%%%%%%%%%%%%%%%%%%%%%%%%%%%%%%%%%%%%%%%%%%%%%%%%%%%%%%%%%%%%%%%%%%%%%%%%%%%%%%%%%%%%%%%%%%
 \begin{table}
 \caption{The values of Wilson coefficients $C_i(m_b)$ in the leading
logarithmic approximation in Standard Model, with
$m_W=80.4\mbox{GeV}$, $m_t=173.8\mbox{GeV}$,
$m_b=4.8\mbox{GeV}$.\cite{Yang:wilsons}}
 \label{tab:wilsons}
 \begin{center}
 \begin{tabular}{c c c c c c c c c}
 \hline\hline
 \ \ \ $C_1$ &$C_2$ &$C_3$ &$C_4$ &$C_5$ &$C_6$ &$C_7$ &$C_9$ &$C_{10}$       \\
 \ \ \ $1.119$   &$-0.270$   &$0.013$    &$-0.027$    &$0.009$    &$-0.033$    &$-0.322$    &$4.344$    &$-4.669$    \\
 \hline\hline
 \end{tabular}
 \end{center}
 \end{table}

 %========================================================================================
 \section{numerical results and discussion}
 \label{section:results}

 \subsection{Form Factors}
 %=========================================================================================
%%%%%%%%%%%%%%%%%%%%%%%%%%%%%%%%%%%%%%%%%%%%%%%%%%%%%%%%%%%%%%%%%%%%%%%%%%%%%%%%%%%%%%%%%%%%%%%%%%%%%%%%%%%%%%%%
%%%%%%%%%%%%%%%%%%%%%%%%%% data in scenario 1 %%%%%%%%%%%%%%%%%%%%%%%%%%%%%%%%%%%%%%%%%%%%%%%%%%%%%%%%%%%%%%%%%%
%%%%%%%%%%%%%%%%%%%%%%%%%%%%%%%%%%%%%%%%%%%%%%%%%%%%%%%%%%%%%%%%%%%%%%%%%%%%%%%%%%%%%%%%%%%%%%%%%%%%%%%%%%%%%%%%
 \begin{table}
 \caption{Form factors for $B\to S$ in scenario
1. The errors arise from the uncertainties of hadronic parameters of
$B_{(s)}$ meson($f_b$ and $\omega_b$), $\Lambda_{\rm{QCD}}$,
scales($t_e^i)$ and the Gegenbauer moments of scalar mesons.
}\label{tab:resultsscenario1}
 \begin{center}
 \begin{tabular}{|c|c c c c c c c c|}
 \hline
 \ \ \        &$F_0(0)=F_1(0)$  &$F_T(0)$   &$a(F_0)$  &$b(F_0)$    &$a(F_1)$    &$b(F_1)$ &$a(F_T)$ &$b(F_T)$  \\
 \hline
 \ \ \ $B\to \sigma$    &$0.28_{-0.06}^{+0.07}$   &$0.29_{-0.06}^{+0.07}$  &$0.65_{-0.07}^{+0.01}$  &$-0.11_{-0.13}^{+0.00}$  &$1.61_{-0.06}^{+0.04}$  &$0.56_{-0.10}^{+0.04}$  &$1.67_{-0.05}^{+0.05}$  &$0.62_{-0.06}^{+0.06}$\\
 \ \ \ $B\to a_0(980)$    &$0.39_{-0.08}^{+0.10}$   &$0.45_{-0.10}^{+0.11}$  &$0.72_{-0.03}^{+0.08}$  &$-0.16_{-0.00}^{+0.12}$  &$1.68_{-0.06}^{+0.03}$  &$0.62_{-0.10}^{+0.01}$  &$1.70_{-0.03}^{+0.06}$  &$0.63_{-0.01}^{+0.11}$\\
 \ \ \ $B\to \kappa(800)$ &$0.27_{-0.06}^{+0.07}$   &$0.29_{-0.07}^{+0.07}$  &$0.71_{-0.08}^{+0.04}$  &$-0.12_{-0.12}^{+0.02}$  &$1.65_{-0.04}^{+0.06}$  &$0.59_{-0.04}^{+0.08}$  &$1.69_{-0.05}^{+0.06}$  &$0.65_{-0.06}^{+0.08}$\\
 \hline\hline
 \ \ \ $B\to f_0(1370)$   &$-0.30_{-0.09}^{+0.08}$   &$-0.39_{-0.11}^{+0.10}$  &$0.70_{-0.02}^{+0.07}$  &$-0.24_{-0.05}^{+0.15}$  &$1.63_{-0.05}^{+0.09}$  &$0.53_{-0.08}^{+0.14}$  &$1.60_{-0.04}^{+0.06}$  &$0.50_{-0.05}^{+0.08}$\\
 \ \ \ $B\to a_0(1450)$   &$-0.31_{-0.09}^{+0.08}$   &$-0.41_{-0.12}^{+0.10}$  &$0.70_{-0.02}^{+0.13}$  &$-0.26_{-0.00}^{+0.24}$  &$1.63_{-0.04}^{+0.08}$  &$0.53_{-0.06}^{+0.13}$  &$1.62_{-0.07}^{+0.04}$  &$0.54_{-0.13}^{+0.03}$\\
 \ \ \ $B\to K^*_0(1430)$ &$-0.34_{-0.09}^{+0.07}$   &$-0.44_{-0.11}^{+0.10}$  &$0.72_{-0.04}^{+0.04}$  &$-0.18_{-0.05}^{+0.04}$  &$1.65_{-0.07}^{+0.04}$  &$0.57_{-0.14}^{+0.08}$  &$1.61_{-0.05}^{+0.04}$  &$0.52_{-0.06}^{+0.05}$\\
 \hline\hline
 \ \ \ $\bar B_s^0\to f_0(980)$     &$0.35_{-0.07}^{+0.09}$   &$0.40_{-0.08}^{+0.10}$   &$0.73_{-0.06}^{+0.04}$   &$-0.13_{-0.09}^{+0.02}$   &$1.66_{-0.05}^{+0.06}$   &$0.60_{-0.05}^{+0.07}$   &$1.70_{-0.04}^{+0.06}$  &$0.66_{-0.05}^{+0.06}$  \\
 \ \ \ $\bar B_s^0\to \kappa(800)$  &$0.29_{-0.06}^{+0.07}$  &$0.31_{-0.06}^{+0.07}$  &$0.66_{-0.03}^{+0.07}$  &$-0.17_{-0.00}^{+0.11}$  &$1.62_{-0.05}^{+0.03}$  &$0.56_{-0.09}^{+0.00}$  &$1.68_{-0.03}^{+0.05}$  &$0.62_{-0.01}^{+0.10}$\\
 \ \ \ $\bar B_s^0\to f_0(1500)$    &$-0.26_{-0.08}^{+0.09}$  &$-0.34_{-0.10}^{+0.10}$  &$0.72_{-0.08}^{+0.14}$  &$-0.20_{-0.10}^{+0.10}$  &$1.61_{-0.03}^{+0.13}$  &$0.48_{-0.02}^{+0.27}$  &$1.60_{-0.04}^{+0.06}$   &$0.48_{-0.04}^{+0.09}$\\
 \ \ \ $\bar B_s^0\to K^*_0(1430)$  &$-0.32_{-0.07}^{+0.06}$  &$-0.41_{-0.09}^{+0.08}$  &$0.69_{-0.03}^{+0.05}$  &$-0.21_{-0.03}^{+0.11}$  &$1.62_{-0.03}^{+0.06}$  &$0.52_{-0.04}^{+0.14}$  &$1.62_{-0.06}^{+0.01}$  &$0.56_{-0.16}^{+0.00}$\\
 \hline
 \end{tabular}
 \end{center}
 \end{table}

 \begin{table}
 \caption{Form factors for $B\to S$ in scenario
 2, with the same error sources as the data in Table \ref{tab:resultsscenario1}.}\label{tab:resultsscenario2}
 \begin{center}
 \begin{tabular}{|c|c c c c c c c c|}
 \hline
 \ \ \        &$F_0(0)=F_1(0)$  &$F_T(0)$   &$a(F_0)$  &$b(F_0)$    &$a(F_1)$    &$b(F_1)$ &$a(F_T)$ &$b(F_T)$  \\
 \hline
 \ \ \ $B\to f_0(1370)$       &$0.63_{-0.14}^{+0.23}$   &$0.76_{-0.17}^{+0.37}$  &$0.70_{-0.11}^{+0.05}$  &$-0.14_{-0.09}^{+0.02}$  &$1.60_{-0.05}^{+0.15}$  &$0.53_{-0.09}^{+0.18}$  &$1.63_{-0.05}^{+0.07}$  &$0.57_{-0.07}^{+0.07}$   \\
 \ \ \ $B\to a_0(1450)$       &$0.68_{-0.15}^{+0.19}$   &$0.92_{-0.21}^{+0.30}$  &$0.62_{-0.08}^{+0.05}$  &$-0.21_{-0.02}^{+0.06}$  &$1.73_{-0.07}^{+0.12}$  &$0.70_{-0.11}^{+0.16}$  &$1.68_{-0.04}^{+0.06}$  &$0.61_{-0.02}^{+0.10}$   \\
 \ \ \ $B\to K^{*}_0(1430)$   &$0.60_{-0.15}^{+0.18}$   &$0.78_{-0.19}^{+0.25}$  &$0.68_{-0.05}^{+0.07}$  &$-0.18_{-0.01}^{+0.06}$  &$1.70_{-0.07}^{+0.09}$  &$0.65_{-0.10}^{+0.10}$  &$1.68_{-0.04}^{+0.07}$  &$0.61_{-0.02}^{+0.11}$   \\
 \hline\hline
 \ \ \ $\bar B_s^0\to f_0(1500)$      &$0.60_{-0.12}^{+0.20}$  &$0.82_{-0.16}^{+0.30}$  &$0.65_{-0.10}^{+0.04}$  &$-0.22_{-0.02}^{+0.07}$  &$1.76_{-0.08}^{+0.13}$  &$0.71_{-0.08}^{+0.20}$  &$1.71_{-0.07}^{+0.04}$   &$0.66_{-0.10}^{+0.06}$\\
 \ \ \ $\bar B_s^0\to K^{*}_0(1430)$  &$0.56_{-0.13}^{+0.16}$  &$0.72_{-0.17}^{+0.22}$  &$0.67_{-0.07}^{+0.06}$  &$-0.17_{-0.07}^{+0.01}$  &$1.69_{-0.07}^{+0.08}$  &$0.63_{-0.10}^{+0.09}$  &$1.68_{-0.06}^{+0.06}$  &$0.63_{-0.08}^{+0.07}$\\
 \hline
 \end{tabular}
 \end{center}
 \end{table}

Our results of  the $B\to S$ form factors are listed in table
\ref{tab:resultsscenario1} and \ref{tab:resultsscenario2}. The
errors for the form factors in those two tables arise from the
uncertainties of hadronic parameters of $B_{(s)}$ meson($f_B$ and
$\omega_b$), $\Lambda_{\rm{QCD}}$($0.20\rm{GeV}$-$0.30\rm{GeV}$),
factorization scales(see Eqs.(\ref{eq:scales})) and the Gegenbauer
moments of scalar mesons. A number of remarks will be given in
order.
\begin{itemize}
\item Compared with transitions of B
meson to pseudoscalar mesons, vector mesons and axial-vector mesons
\cite{Lu:Btolight,formfactors}, our predictions on $B\to S$ form
factors in scenario 2 are obviously larger, which is caused mainly
by the large decay constants($\bar f_S$) of the scalar mesons. For
example, the form factor $F_0(0)$ of B meson to pion transition is
about $0.23$\cite{Lu:Bs} with $0.131\rm{GeV}$ as the decay constant
of pion, while the B meson to $a_0(980)$ transition in scenario 1
has $0.39$ as its corresponding form factor, whose decay constant is
more than two times larger than pion.

\item In Table \ref{tab:resultsscenario1}, the form factors of $B\to
\sigma$ are smaller than those of $B\to a_0(980)$. Because the same
decay constant and Gegenbauer moments for these two particles are
used in the calculation, the differences are caused by the mass
differences between $a_0(980)$ and $\sigma$($0.98\rm{GeV}$ for
$a_0(980)$ and $0.513\rm{GeV}$ for $\sigma$\cite{pdg}).  In scenario
1, there are small differences between $\kappa(800)$ and $f_0(600)$
in masses($0.672\rm{GeV}$ for $\kappa(800)$), decay constants and
Gegenbauer moments. Besides, the contribution from twist-2 LCDA of
$\kappa(800)$, which is proportional to $f_S$ , is too small to give
sizable differences. Thus the $B\to \sigma$ and $B\to\kappa(800)$
have nearly the same form factors as shown in Table
\ref{tab:resultsscenario1}. Comparing the form factors of $B\to
\kappa(800)$ with $\bar B_s^0\to\kappa(800)$ in Table
\ref{tab:resultsscenario1}, one can find that the differences
between $B$ and $\bar B_s^0$ mesons affect little. Therefore, the
large differences between the form factors of $\bar
B_s^0\to\kappa(800)$ and those of $B\to a_0(980)$ are mainly due to
the large difference between the scalar meson masses.

\item The form factors of B to heavier nonet transition in scenario 1
are negative, while the others are positive. The reason is that the
decay constants($\bar f_S$) of the heavier nonet in scenario 1 have
opposite signs to the others, which is clearly shown in Table
\ref{tab:Gmoments} and \ref{tab:Gmoments2}.

\item As we can see from the table~\ref{tab:resultsscenario1} and \ref{tab:resultsscenario2},
the predictions in scenario 2 are larger than the corresponding ones
in scenario 1 roughly by a factor of 2 in magnitude. In order to
show how these large differences are generated, we take the form
factor $F_0(0)$ as an example and list contributions from different
terms in LCDAs in Table \ref{tab:Genbauers}(Data is given with
asymptotic forms of twist-3 LCDAs are adopted in both scenario 1 and
scenario 2, because the terms with Gegenbauer moments bring so small
effects, which is discussed in the following, that they can't change
the argument). The contributions from the two twist-3 LCDAs
$\phi_S^s$ and $\phi_S^T$ are given in the first two columns. The
numbers in the column '$B_1$' denotes the contributions from the
Gegenbauer moments $B_1$ in twist-2 LCDAs. It is also similar for
the fourth $B_3$ column. The last column collects the total
contributions to the form factors. The different inputs between in
scenario 1 and in scenario 2 are the decay constants and Gegenbauer
moments. If only twist-3 LCDAs are taken into account, the form
factors will be proportional to the decay constant. Since the decay
constants $\bar f_S$ in S2 are (typically 60\%) larger than those in
S1 in magnitude,  the form factors are accordingly larger. The
$\phi_S^s$ term give much larger contributions than the $\phi_S^T$
term. Contributions from the Gegenbauer moments of the twist-2 LCDAs
sizably enhance the form factors in S2 but not too much in S1. For
$B$ to scalar meson transitions in scenario 1, the $B_1$ terms
provide contributions with the same sign with the twist-3 terms,
while the terms with $B_3$ have the opposite sign. Thus the two
terms of the twist-2 LCDAs give destructive contributions to the
total form factors in S1. The situation is different in S2, although
the two Gegenbauer moments are small in magnitude, they give
constructive contributions and induce much larger form factors.

\item We also investigate the contributions from terms with
Gegenbauer moments in twist-3 LCDAs, and find that the effects
brought by these moments are not large. Taking $B\to f_0(1370)$
transition as an example, a comparison between the cases with and
without these contributions is given in Table \ref{tab:twist3}. We
can see that most of the results are changed by less than $10\%$.

\item Compared with our previous study on $B\to f_0, K_0^*(1430)$
transitions~\cite{Wang:2006ria,Shen:2006ms}, the predictions for the
form factors given in the present work are a bit smaller. The main
reason is that different values for the threshold resummation
parameters $c$ have been used. Moreover, the form factors in this
paper are larger than those obtained in other approaches or
models\cite{results:LFQM,results:QCDSR,MZyang,YMwang}. As a result,
the branching ratios of the semileptonic decays are larger, which is
discussed in the following.
\end{itemize}

As we have mentioned in the introduction section, the
experimentalists have already provided many investigations on
nonleptonic B decays involving a scalar meson in the final state.
Among these decays, the so-called color-allowed tree-dominated
processes can be directly utilized to estimate the $B\to S$ form
factors, under the hypothesis of factorization. For example, the
$\bar B^0\to a_0^+\pi^-$ decay amplitude in the factorization scheme
is expressed as:
\begin{eqnarray}
 {\cal A}(\bar B^0\to a_0^+\pi^-)&=& \frac{G_F}{\sqrt 2}m_B^2 f_\pi
 F_0^{B\to a_0}\left\{V_{ub}V_{ud}^* [a_1+a_4+a_{10}-r_\pi
 (a_6+a_8)]\right.\nonumber\\
 &&\;\;\;\; \left.+V_{cb}V_{cd}^* [a_4+a_{10}-r_\pi
 (a_6+a_8)]\right\},
\end{eqnarray}
where $a_i$ is the combination of Wilson coefficient
\begin{eqnarray}\label{eq:combination}
  &&a_1= C_2+C_1/3,  \;\;a_2= C_1+C_2/3, \nonumber \\
  &&a_i=C_i+C_{i+1}/N_c\;\; (i=3,5,7,9),\nonumber \\
  &&a_i=C_i+C_{i-1}/N_c\;\; (i=4,6,8,10).
\end{eqnarray}
$a_1\sim 1$, and it has small uncertainties. Although there are
large uncertainties for $a_3$-$a_{10}$, the combination of Wilson
coefficients satisfies:
\begin{eqnarray}
 a_1\gg {\rm max}[a_{3-10}].
\end{eqnarray}
If only the branching ratios are concerned, contributions from the
penguin operators ($a_{3-10}$ terms) can be safely neglected and
thus
\begin{eqnarray}
 {\cal A}(\bar B^0\to a_0^+\pi^-)&=& \frac{G_F}{\sqrt 2}m_B^2 f_\pi
 F_0^{B\to a_0} V_{ub}V_{ud}^* a_1.
\end{eqnarray}
If the partial decay widths are well determined experimentally,
these results will directly constrain the B to scalar meson
transition form factors. The upper bounds for $B\to a_0\pi$ are
given as(in unit of $10^{-6}$):
\begin{eqnarray}
 &&{\cal BR} (B \to a_0^\pm(980)\pi^\mp)<3.1,\nonumber\\
 &&{\cal BR} (B \to a_0^\pm(1450)\pi^\mp)<2.3,
\end{eqnarray}
where the daughter BF has taken to be $100\%$. Since the scalar
mesons $a_0(980)$ and $a_0(1450)$ have vanishing decay constants in
the isospin limit, the branching ratios of $\bar B^0\to a_0^-\pi^+$
are very small and one expects the relation: ${\cal BR} (B \to
a_0^\pm\pi^\mp)= {\cal BR} (\bar B^0 \to a_0^+\pi^-)$. Compared with
the branching ratio of $\bar B^0\to \pi^+\pi^-$(in unit of
$10^{-6}$)
\begin{eqnarray}
 &&{\cal BR} (B \to \pi^+\pi^-)=5.16\pm0.22,
\end{eqnarray}
results provide the upper bound for the $B\to a_0$ form factors:
\begin{eqnarray}
 F_0(B\to a_0(980))<0.78 F_0(B\to\pi)=0.18,\;\;\; F_0(B\to
 a_0(1450))<0.67
 F_0(B\to\pi)=0.15,
\end{eqnarray}
where as an rough estimation, we have taken
$F_0(B\to\pi)=0.23$\cite{Lu:Bs}. Compared with our results in Table
\ref{tab:resultsscenario1} and \ref{tab:resultsscenario2}, one can
see our results have exceeded the present experimental upper bound.
Despite of that, it does not mean our predictions are ruled out by
the data, as the daughter decay is not taken into account in the
derivation for the experimental bound. Our predictions will be
confronted with the real bound in the future, whenever the daughter
decay of $a_0$ is well studied.

%%%%%%%%%%%%%%%%%%%%%%%%%%%%%%%%%%%%%%%%%%%%%%%%%%%%%%%%%%%%%%%%%%%%%%%%%%%%%%%%%%%%%%%%%%%%%%%%%%%%%%%%%%%%%%%%%%%%%%%%%%%%%
%%%%%%%%%%%%%%%%%%%%%%%%%%%%%%%%%%%%%%%%%%%%%%%%%%%%%%%%%%%%%%%%%%%%%%%%%%%%%%%%%%%%%%%%%%%%%%%%%%%%%%%%%%%%%%%%%%%%%%%%%%%%%
 \begin{table}
\caption{Contributions from different LCDAs to the $B\to S$ form
factor $F_0$ in scenario 1 (S1) or scenario 2 (S2).  The
contributions from the two twist-3 LCDAs $\phi_S^s$ and $\phi_S^T$
are given in the first two columns.  The numbers in the column'
$B_1$' denotes the contributions from the Gegenbauer moments $B_1$
in twist-2 LCDAs.  It is also similar for the fourth column. The
last column collects the total contributions to the form
factors(Data is given with asymptotic forms of twist-3 LCDAs adopted
in both scenario 1 and scenario 2).}
 \label{tab:Genbauers}
 \begin{center}
 \begin{tabular}{c|c c c c c c}
 \hline\hline
 \ \ \      &     & $\phi_S^s$ & $\phi_S^T$ &$B_1$    &$B_3$      & Total  \\
 \hline
 \ \ \ $B\to a_0(1450)$   &$\rm{S1:}$  &$-0.21$  &$-0.05$  &$0.14$  &$-0.19$ &$-0.31$  \\
 \ \ \                    &$\rm{S2:}$  &$0.35$   &$0.08$   &$0.15$  & $0.11$ &$0.69$  \\
 \hline
 \ \ \ $B\to K^*_0(1430)$   &$\rm{S1:}$  &$-0.22$  &$-0.05$ &$0.10$  &$-0.18$  &$-0.34$  \\
 \ \ \                      &$\rm{S2:}$  &$0.33$   &$0.07$   &$0.14$  &$0.09$    &$0.62$  \\
 \hline
 \ \ \ $\bar B_s^0\to f_0(1500)$   &$\rm{S1:}$  &$-0.17$  &$-0.04$  &$0.11$   & $-0.16$&$-0.26$  \\
 \ \ \                             &$\rm{S2:}$  &$0.32$   &$0.08$   &$0.13$&$0.09$   &$0.61$  \\
 \hline
 \ \ \ $\bar B_s^0\to K^*_0(1430)$   &$\rm{S1:}$  &$-0.19$  &$-0.05$  &$0.09$ &$-0.17$  &$-0.32$  \\
 \ \ \                               &$\rm{S2:}$  &$0.27$   &$0.07$   &$0.14$  &$0.09$ &$0.58$  \\
 \hline\hline
 \end{tabular}
 \end{center}
 \end{table}

%%%%%%%%%%%%%%%%%%%%%%%%%%%%%%%%%%%%%%%%%%%%%%%%%%%%%%%%%%%%%%%%%%%%%%%%%%%%%%%%%%%%%%%%%%%%%%%%%%%%%%%%%%%%%%%%%%%%%%%%%%%%%
%%%%%%%%%%%%%%%%%%%%%%%%%%%%%%%%%%%%%%%%%%%%%%%%%%%%%%%%%%%%%%%%%%%%%%%%%%%%%%%%%%%%%%%%%%%%%%%%%%%%%%%%%%%%%%%%%%%%%%%%%%%%%
 \begin{table}
 \caption{Form factors for $B\to f_0(1370)$. The first line and the second line are the
results with and without contributions from the terms with
Gegenbauer moments in twist-3 LCDAs respectively.}
 \label{tab:twist3}
 \begin{center}
 \begin{tabular}{c c c c c c c c}
 \hline
 \ \ \   $F_0(0)=F_1(0)$  &$F_T(0)$   &$a(F_0)$  &$b(F_0)$    &$a(F_1)$    &$b(F_1)$ &$a(F_T)$ &$b(F_T)$  \\
 \hline
 \ \ \ $0.63_{-0.14}^{+0.23}$   &$0.76_{-0.17}^{+0.37}$  &$0.70_{-0.11}^{+0.05}$  &$-0.14_{-0.09}^{+0.02}$  &$1.60_{-0.05}^{+0.15}$  &$0.53_{-0.09}^{+0.18}$  &$1.63_{-0.05}^{+0.07}$  &$0.57_{-0.07}^{+0.07}$   \\
 \ \ \ $0.67_{-0.14}^{+0.17}$   &$0.83_{-0.18}^{+0.21}$  &$0.71_{-0.07}^{+0.02}$  &$-0.12_{-0.11}^{+0.00}$  &$1.64_{-0.05}^{+0.04}$  &$0.57_{-0.07}^{+0.04}$  &$1.65_{-0.04}^{+0.05}$  &$0.59_{-0.03}^{+0.07}$   \\
 \hline\hline
 \end{tabular}
 \end{center}
 \end{table}
%%%%%%%%%%%%%%%%%%%%%%%%%%%%%%%%%%%%%%%%%%%%%%%%%%%%%%%%%%%%%%%%%%%%%%%%%%%%%%%%%%%%%%%%%%%%%%%%%%%%%%%%%%%%%%%%%%%%%%%%%%%%%%
%%%%%%%%%%%%%%%%%%%%%%%%%%%%%%%%%%%%%%%%%%%%%%%%%%%%%%%%%%%%%%%%%%%%%%%%%%%%%%%%%%%%%%%%%%%%%%%%%%%%%%%%%%%%%%%%%%%%%%%%%%%%%%

%---------------------------------------------------------
\subsection{Decay widths and branching fractions}
%---------------------------------------------------------

\begin{figure}
 \begin{center}
 \includegraphics[scale=0.6]{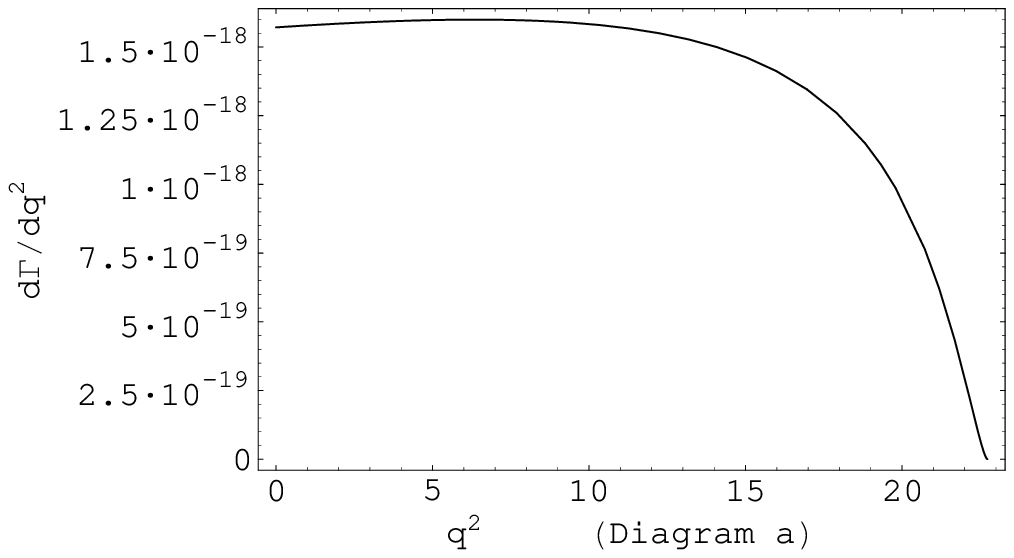}
 \includegraphics[scale=0.6]{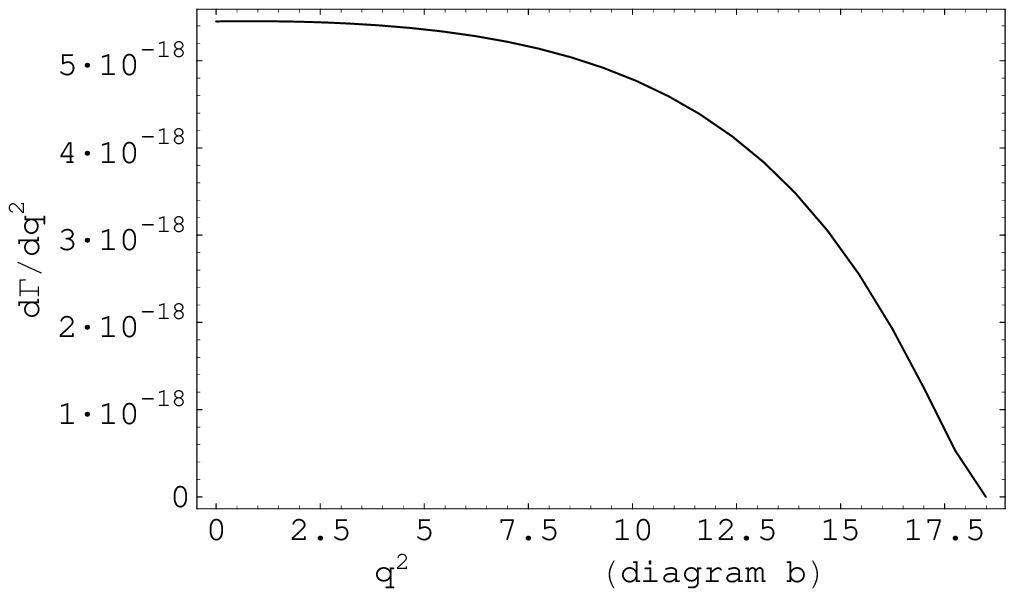}
 \includegraphics[scale=0.6]{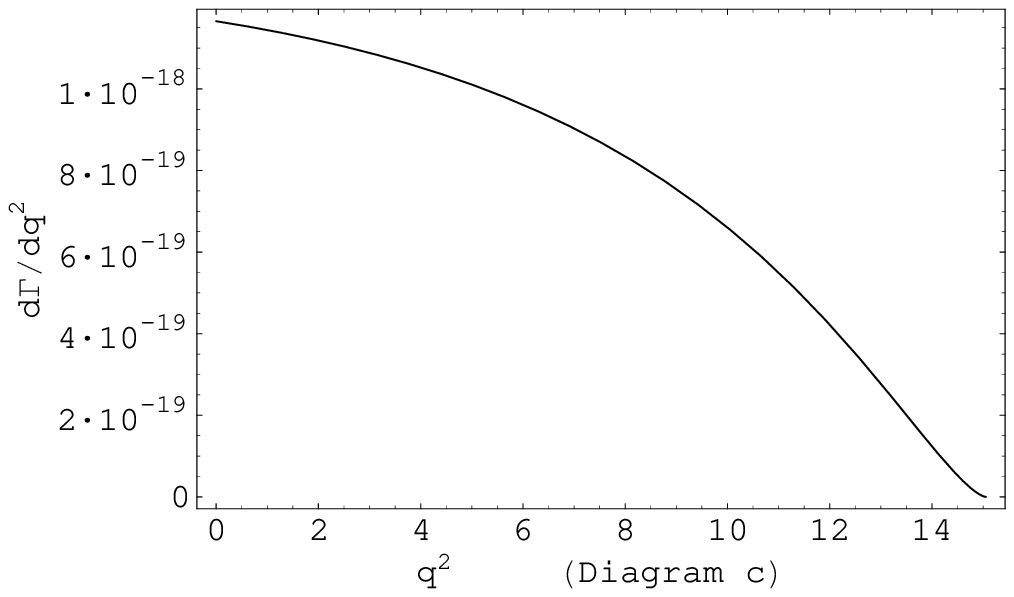}
 \includegraphics[scale=0.6]{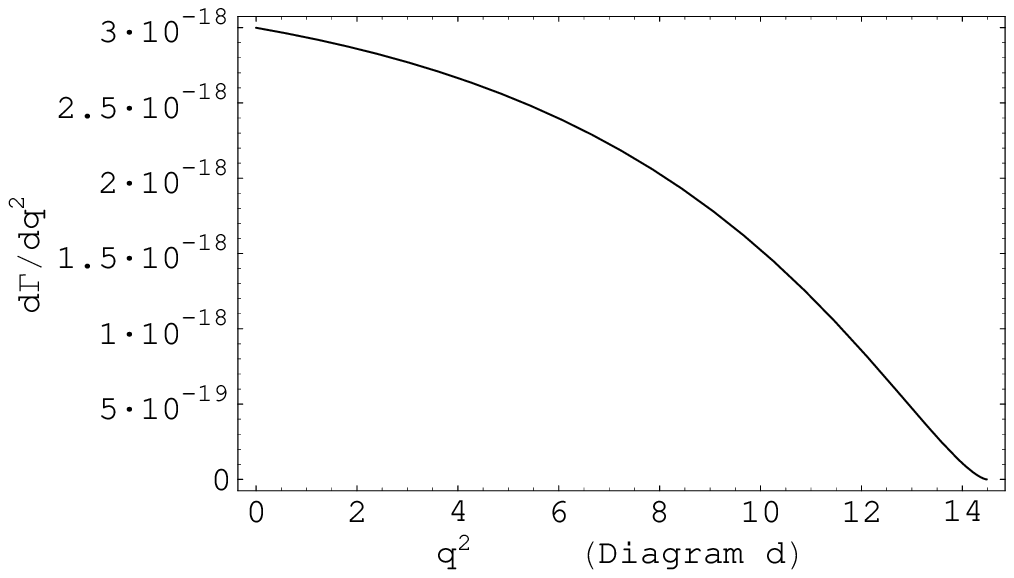}
 \includegraphics[scale=0.6]{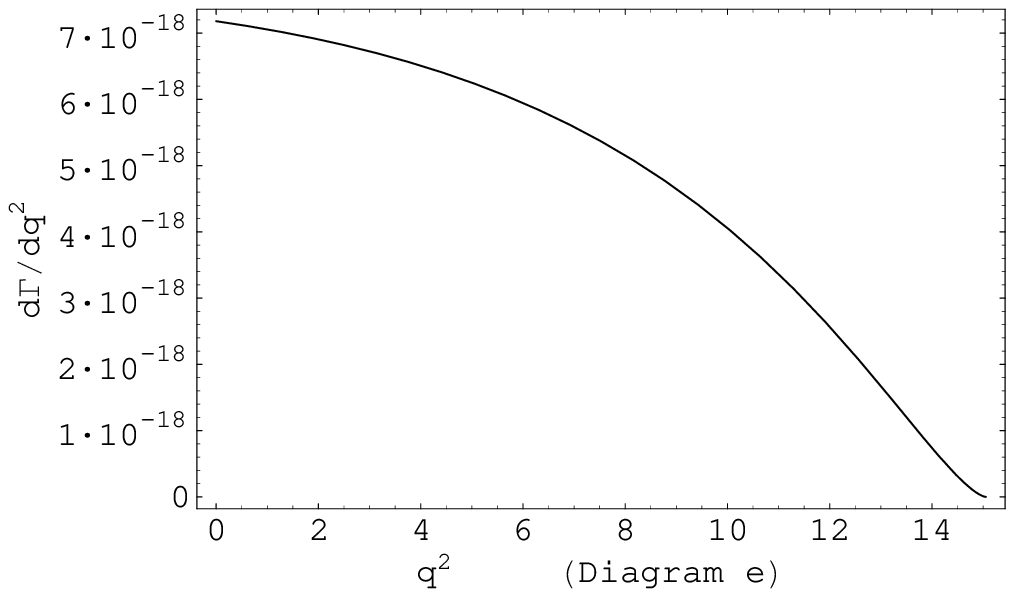}
 \includegraphics[scale=0.6]{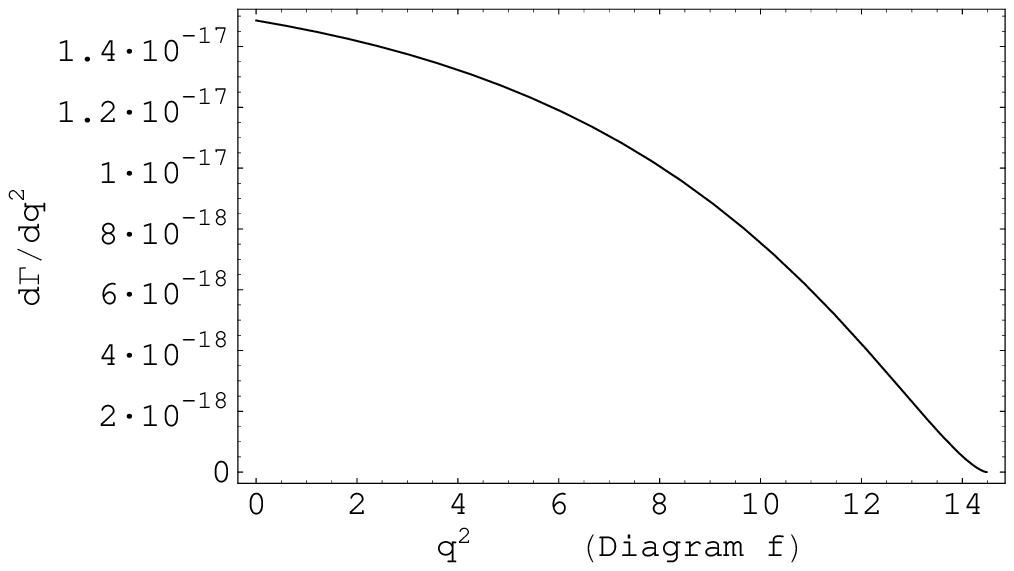}
 \includegraphics[scale=0.6]{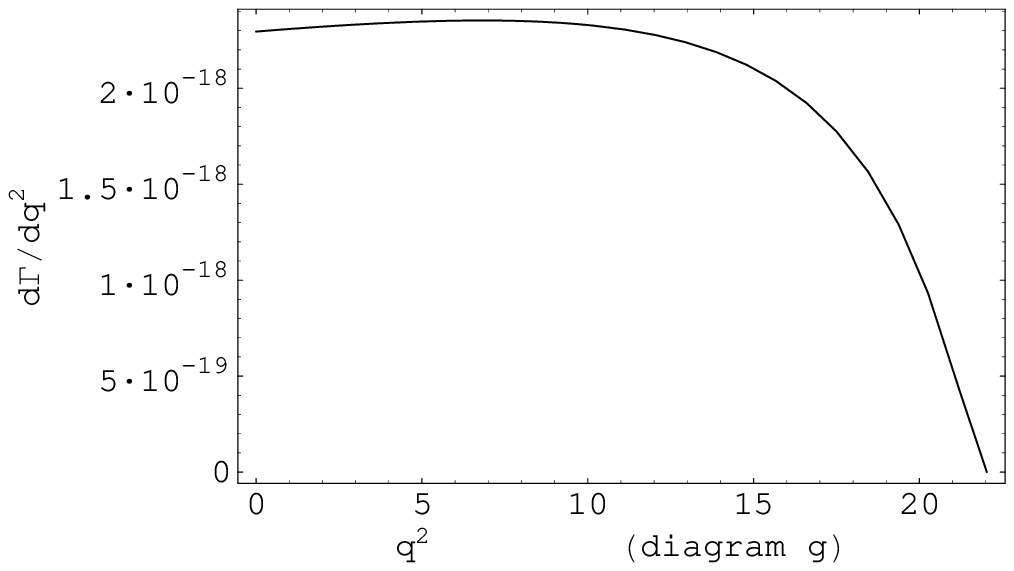}
 \includegraphics[scale=0.6]{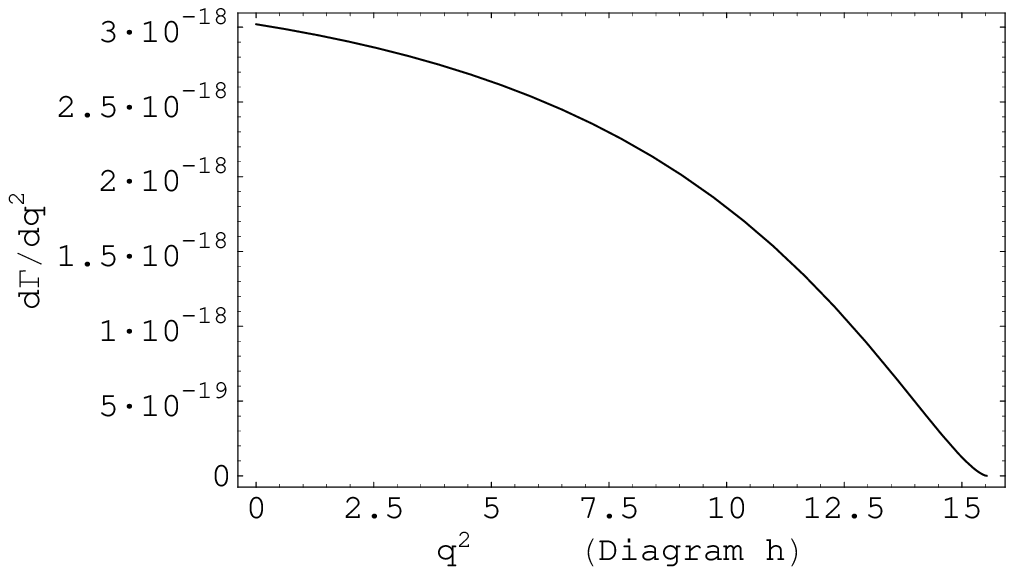}
 \includegraphics[scale=0.6]{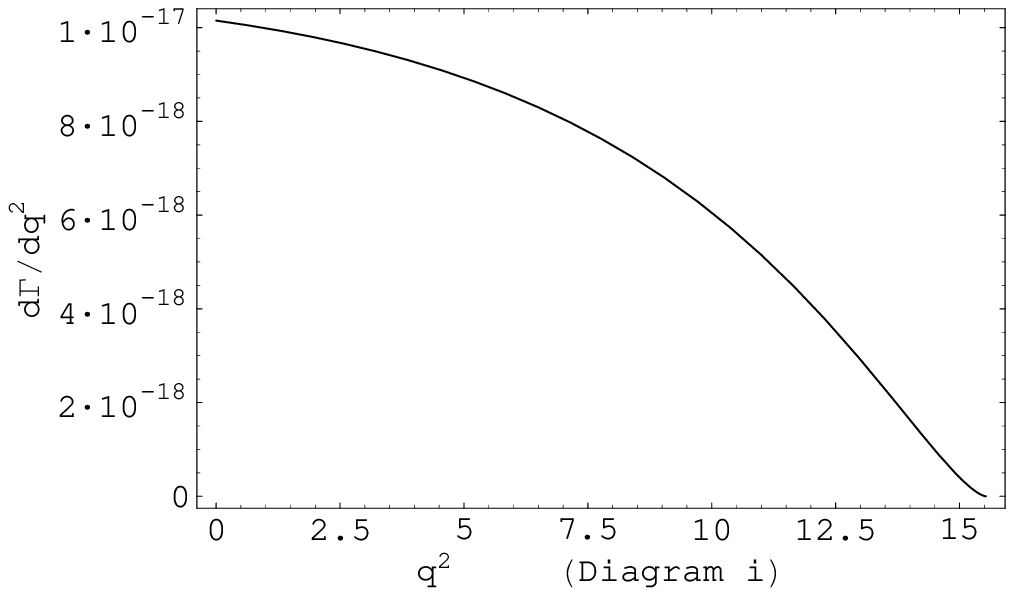}
 \caption{Partial decay widths of the semileptonic  $B\to Sl\bar\nu$ decays as functions of $q^2$.
 Diagram a-d denote the $B^-\to (\sigma,a_0^+(980),f_0(1370),a_0^+(1450)) l^-\bar\nu_l$ in scenario
 1 respectively; Diagram e-f denote the $B^-\to (f_0(1370),a_0^+(1450)) l^-\bar\nu_l$ in scenario
 2 respectively;
 Diagram g: $\bar B_s\to \kappa^+(800) l^-\bar\nu_l$ in scenario 1;
 Diagram h: $\bar B_s\to K^{*+}_0(1430) l^-\bar\nu_l$ in scenario 1;
 Diagram i: $\bar B_s\to K^{*+}_0(1430) l^-\bar\nu_l$ in scenario 2.  }
 \label{fig:gamma1}
 \end{center}
 \end{figure}

\begin{figure}
 \begin{center}
 \includegraphics[scale=0.6]{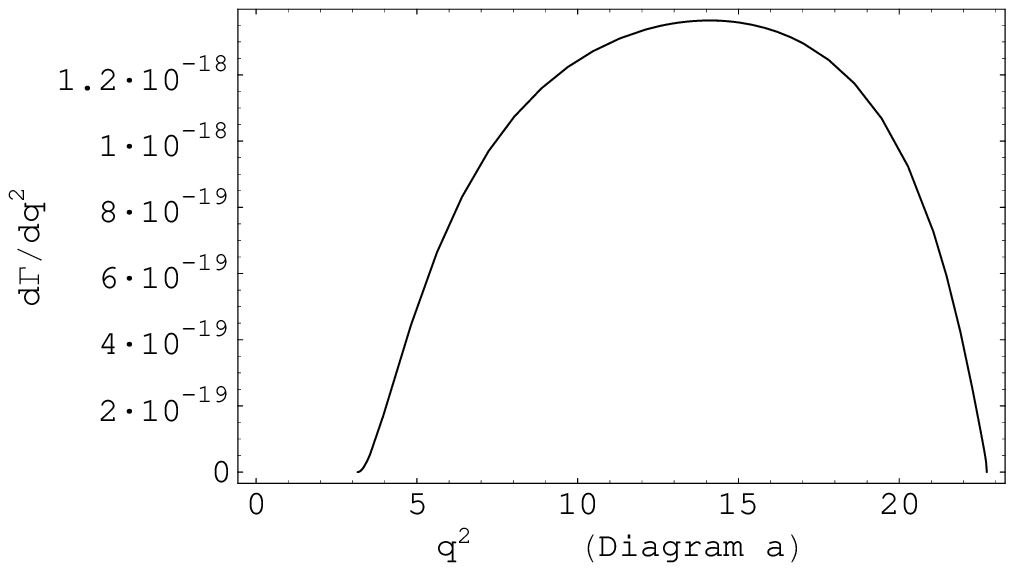}
 \includegraphics[scale=0.6]{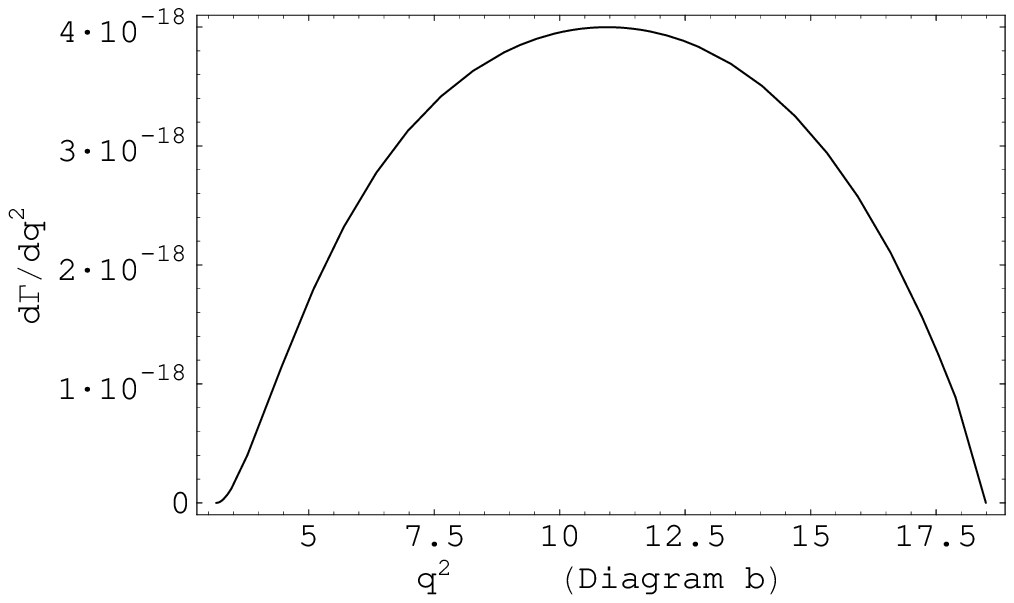}
 \includegraphics[scale=0.6]{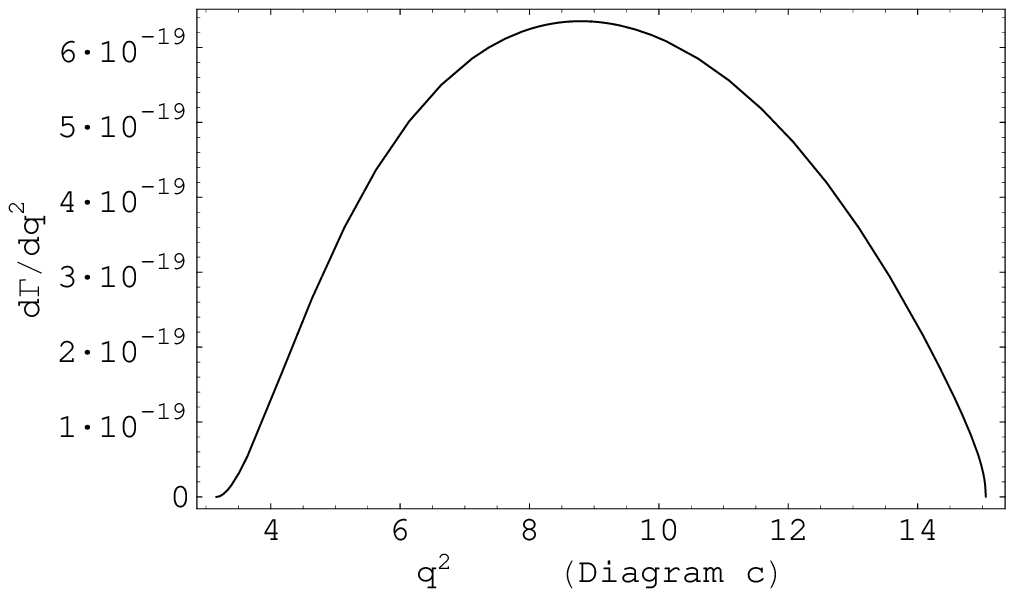}
 \includegraphics[scale=0.6]{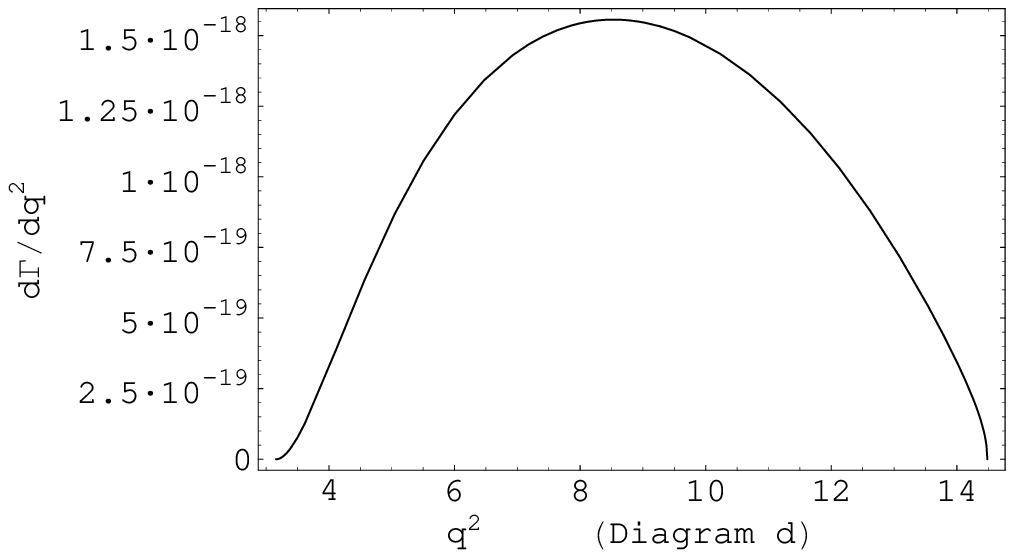}
 \includegraphics[scale=0.6]{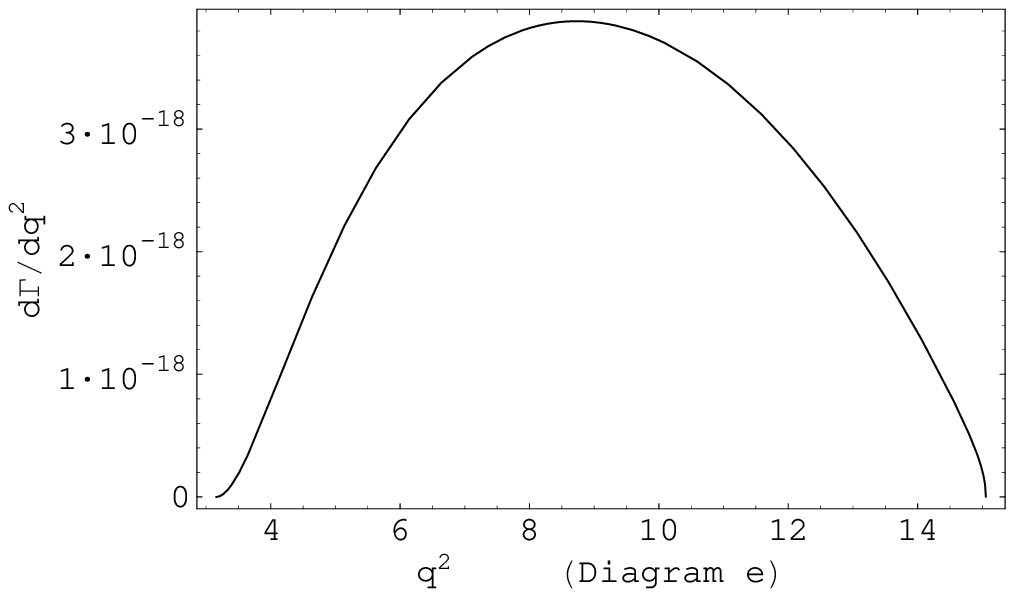}
 \includegraphics[scale=0.6]{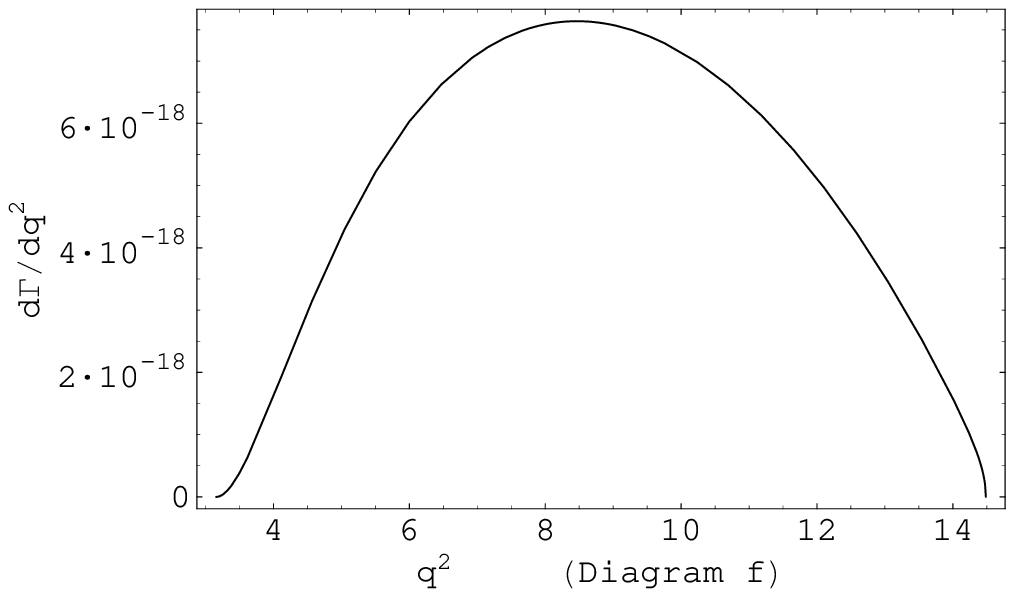}
 \includegraphics[scale=0.6]{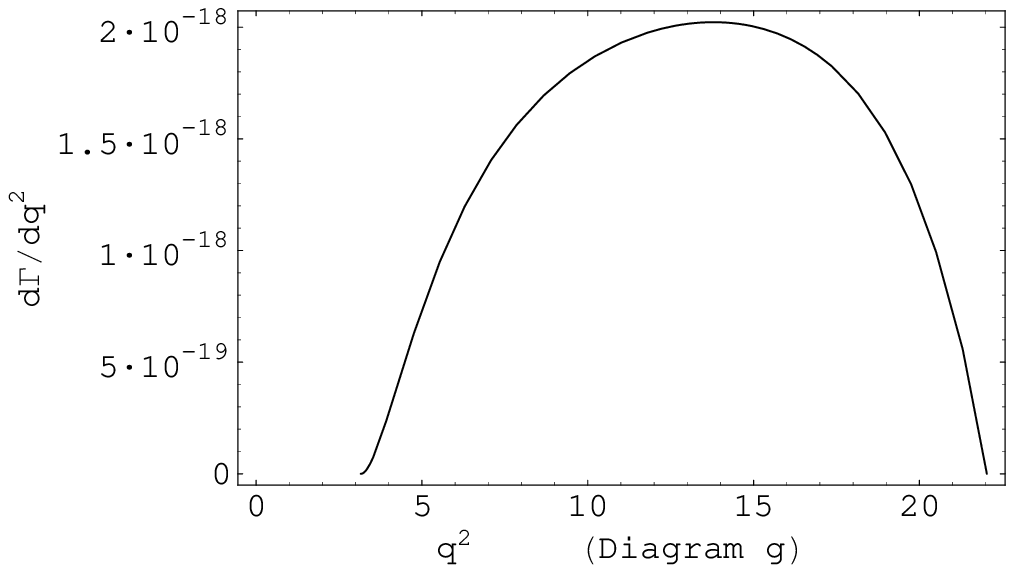}
 \includegraphics[scale=0.6]{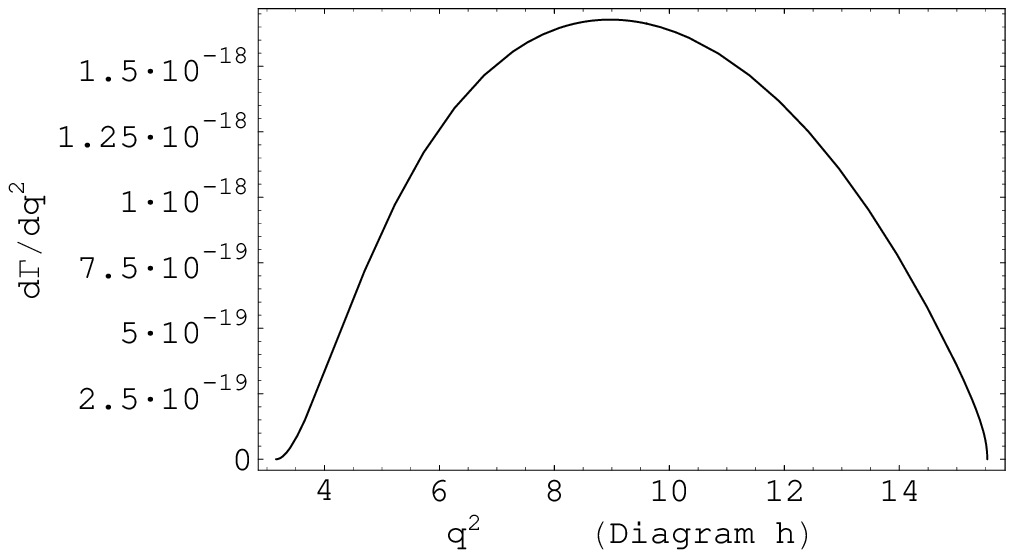}
 \includegraphics[scale=0.6]{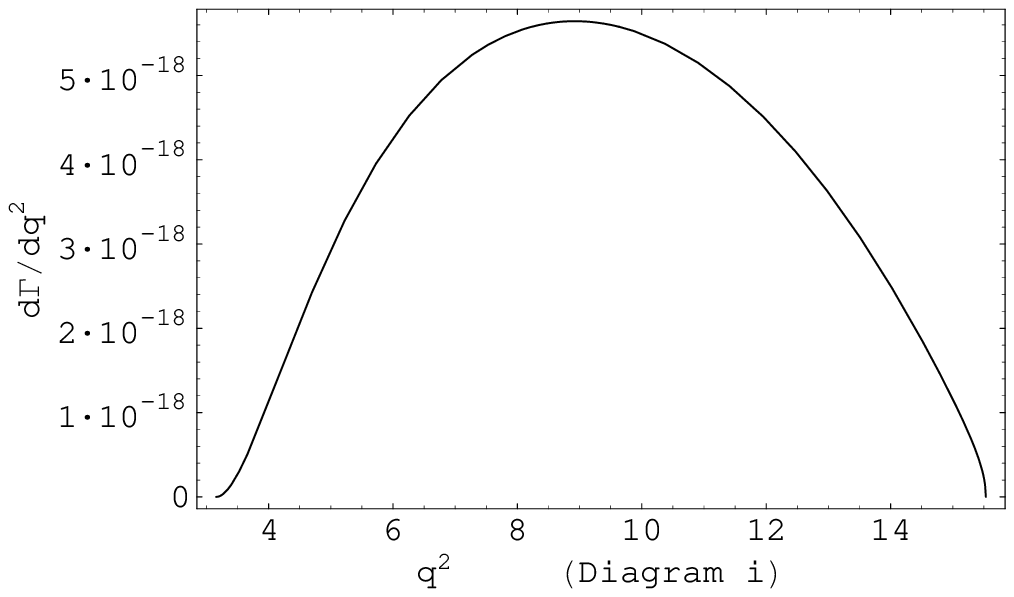}
 \caption{Partial decay widths of the semileptonic  $B\to S\tau\bar\nu$ decays as functions of $q^2$.
 Diagram a: $B^-\to \sigma \tau^-\bar\nu_l$ in scenario 1;
 Diagram b: $\bar B^0\to a_0^+(980) \tau^-\bar\nu_l$ in scenario 1;
 Diagram c: $B^-\to f_0(1370) \tau^-\bar\nu_l$ in scenario 1;
 Diagram d: $\bar B^0\to a_0^+(1450) \tau^-\bar\nu_l$ in scenario 1;
 Diagram e: $B^-\to f_0(1370) \tau^-\bar\nu_l$ in scenario 2;
 Diagram f: $\bar B^0\to a_0^+(1450) \tau^-\bar\nu_l$ in scenario 2;
 Diagram g: $\bar B_s\to \kappa^+(800) \tau^-\bar\nu_l$ in scenario 1;
 Diagram h: $\bar B_s\to K^{*+}_0(1430) \tau^-\bar\nu_l$ in scenario 1;
 Diagram i: $\bar B_s\to K^{*+}_0(1430) \tau^-\bar\nu_l$ in scenario 2.  }
 \label{fig:gamma2}
 \end{center}
 \end{figure}

\begin{figure}
 \begin{center}
 \includegraphics[scale=0.6]{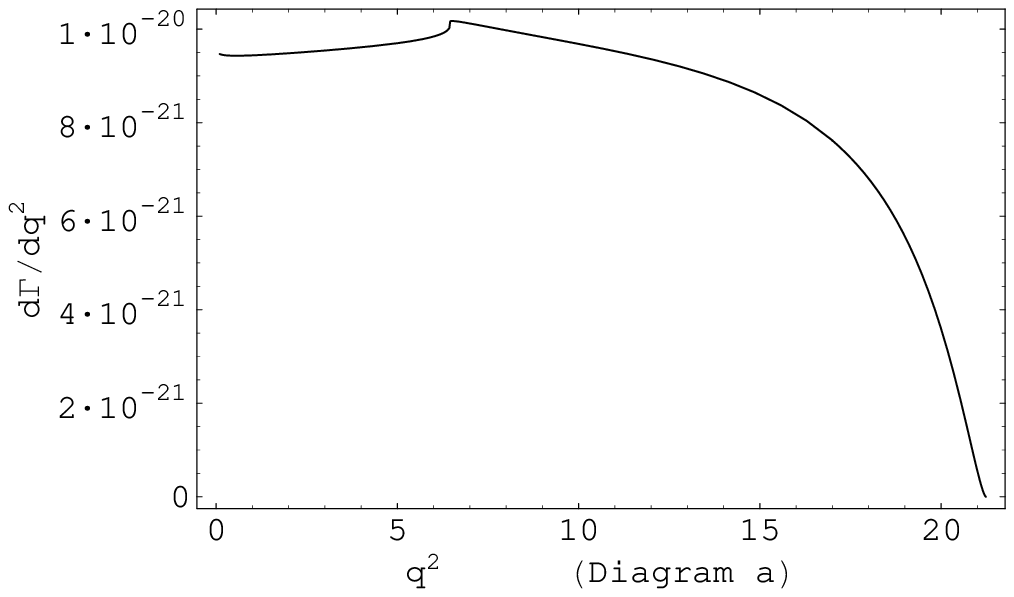}
 \includegraphics[scale=0.6]{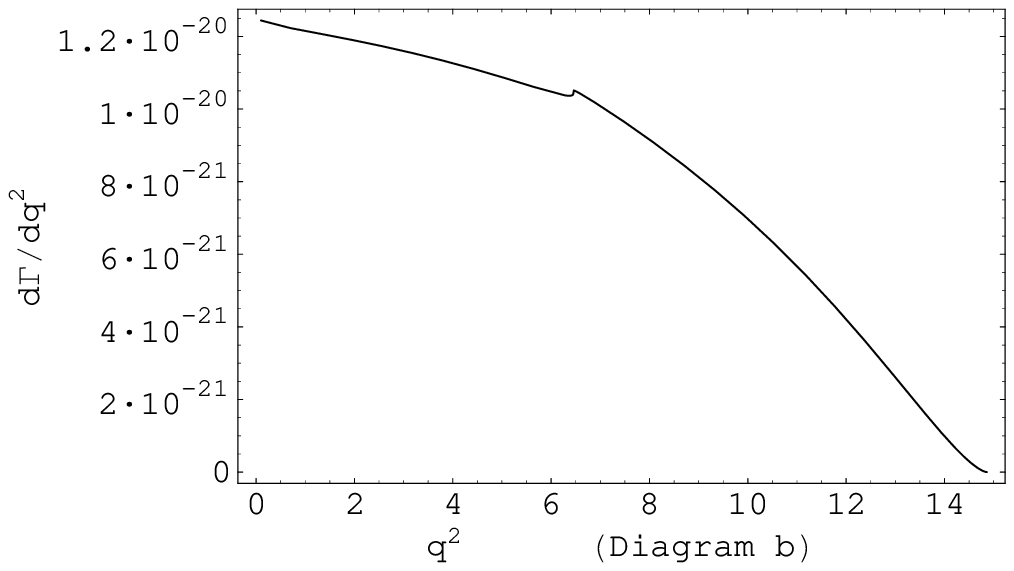}
 \includegraphics[scale=0.6]{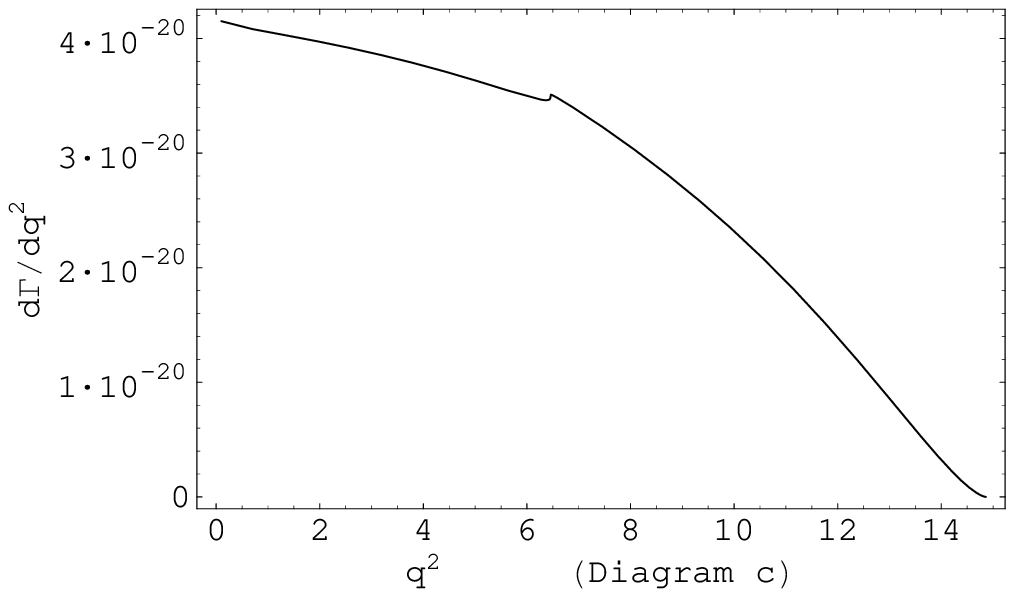}
 \includegraphics[scale=0.6]{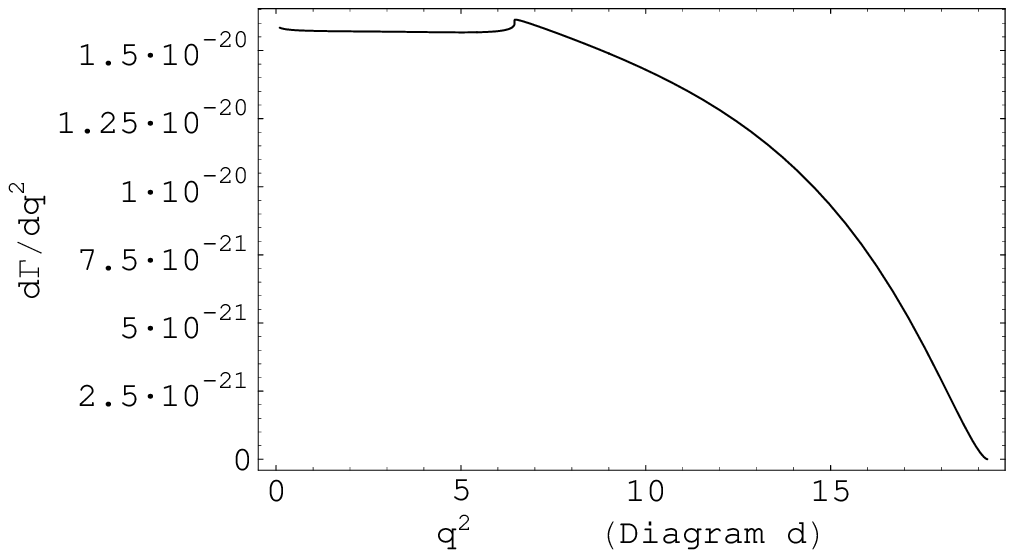}
 \includegraphics[scale=0.6]{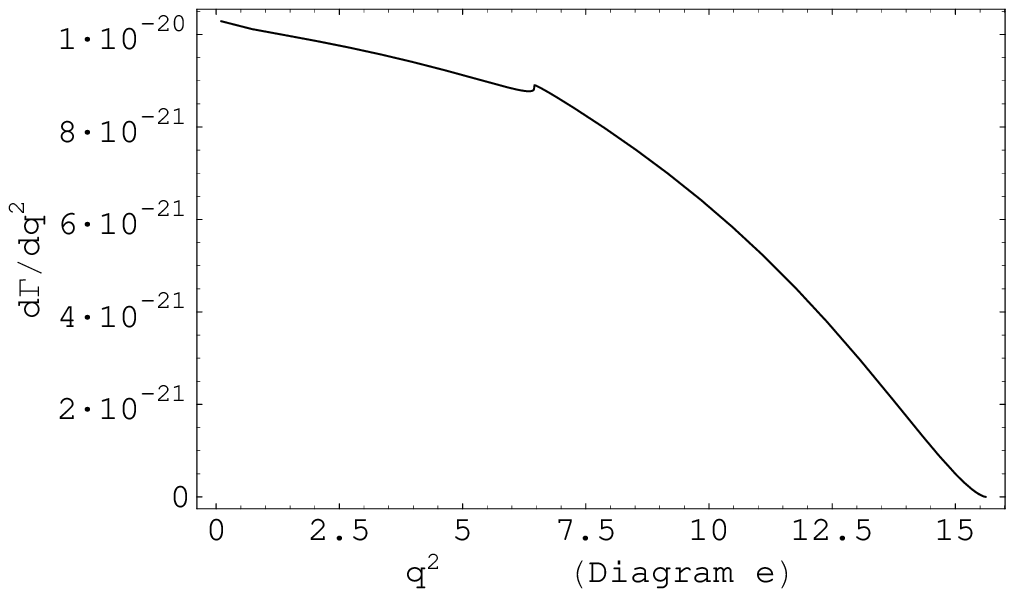}
 \includegraphics[scale=0.6]{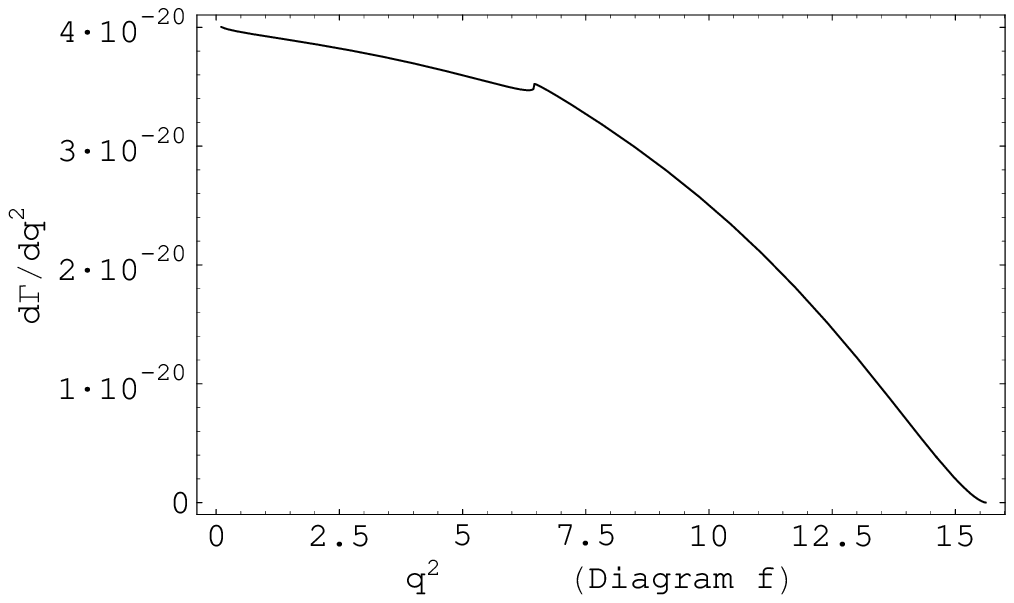}
 \caption{Partial decay widths of the semileptonic  $B\to Sl^+l^-$($l=e$, $\mu$) decays as functions of $q^2$.
 Diagram a: $B^-\to\kappa^-l^+l^-$ in scenario 1;
 Diagram b: $B^-\to K^{*-}_0(1430)l^+l^-$ in scenario 1;
 Diagram c: $B^-\to K^{*-}_0(1430)l^+l^-$ in scenario 2;
 Diagram d: $\bar B_s^0\to f_0(980)l^+l^-$ in scenario 1;
 Diagram e: $\bar B_s^0\to f_0(1500)l^+l^-$ in scenario 1;
 Diagram f: $\bar B_s^0\to f_0(1500)l^+l^-$ in scenario 2;}
 \label{fig:gamma3}
 \end{center}
 \end{figure}

\begin{figure}
 \begin{center}
 \includegraphics[scale=0.6]{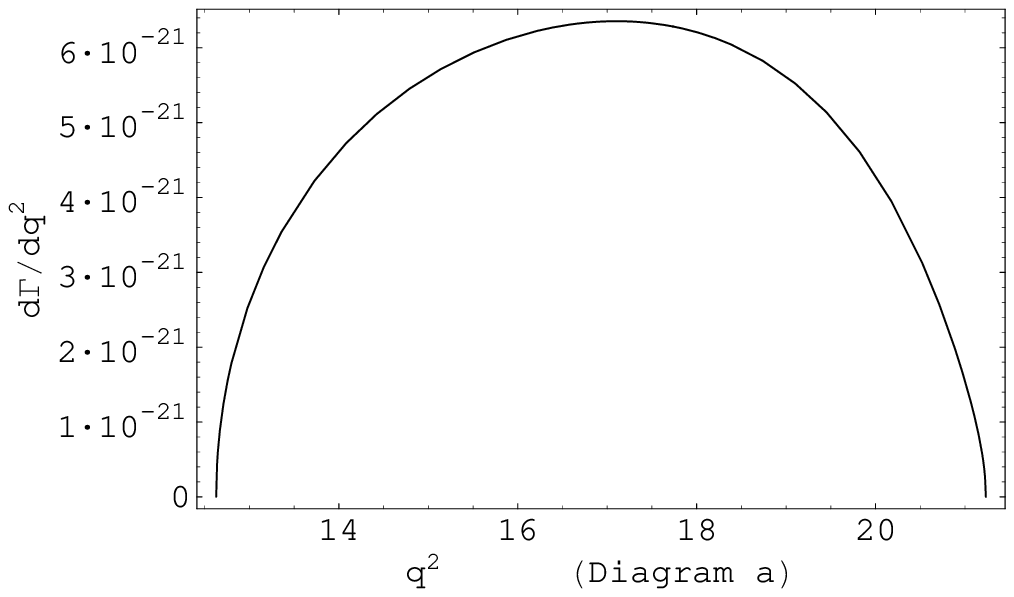}
 \includegraphics[scale=0.6]{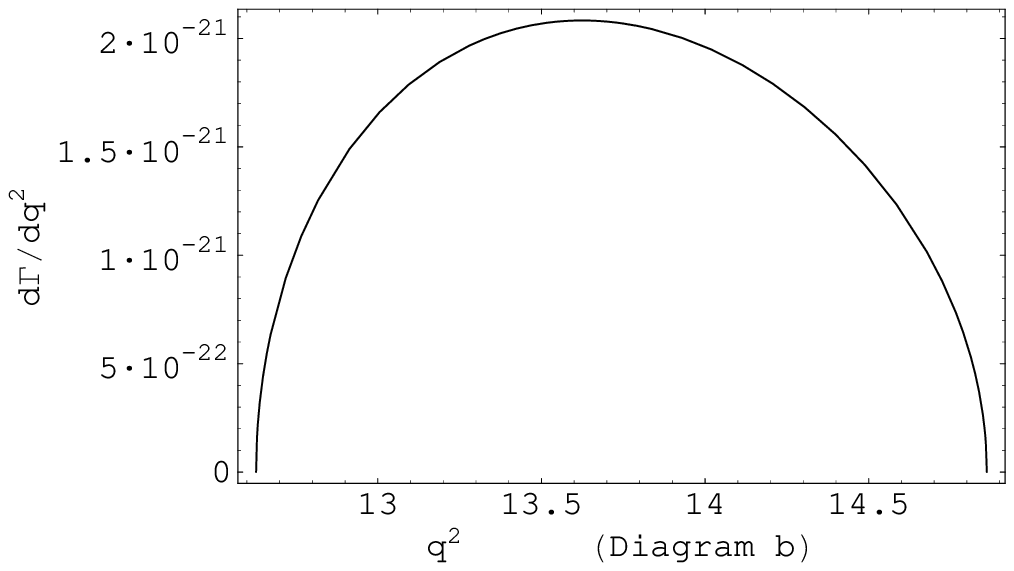}
 \includegraphics[scale=0.6]{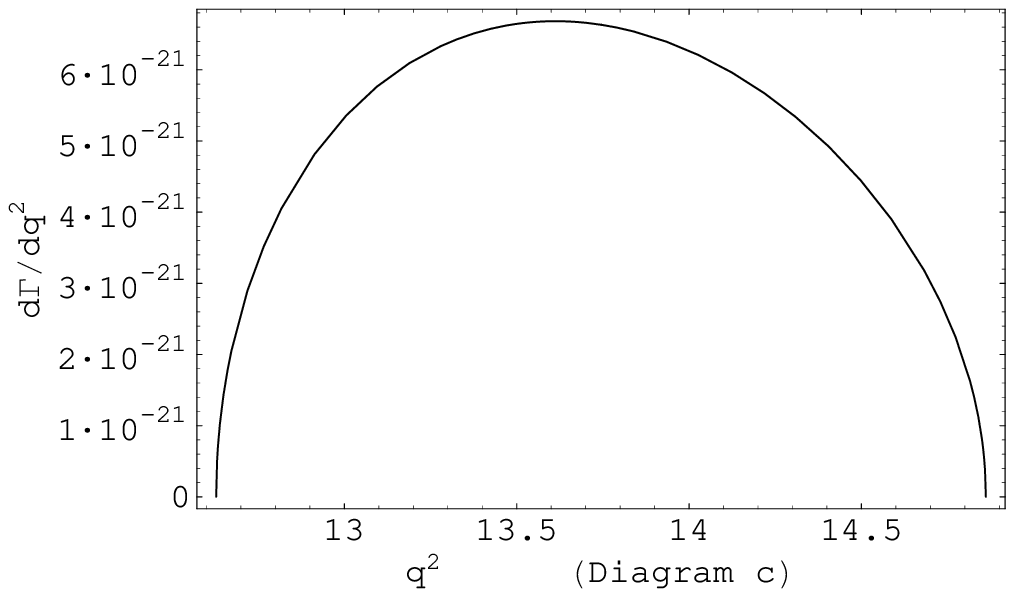}
 \includegraphics[scale=0.6]{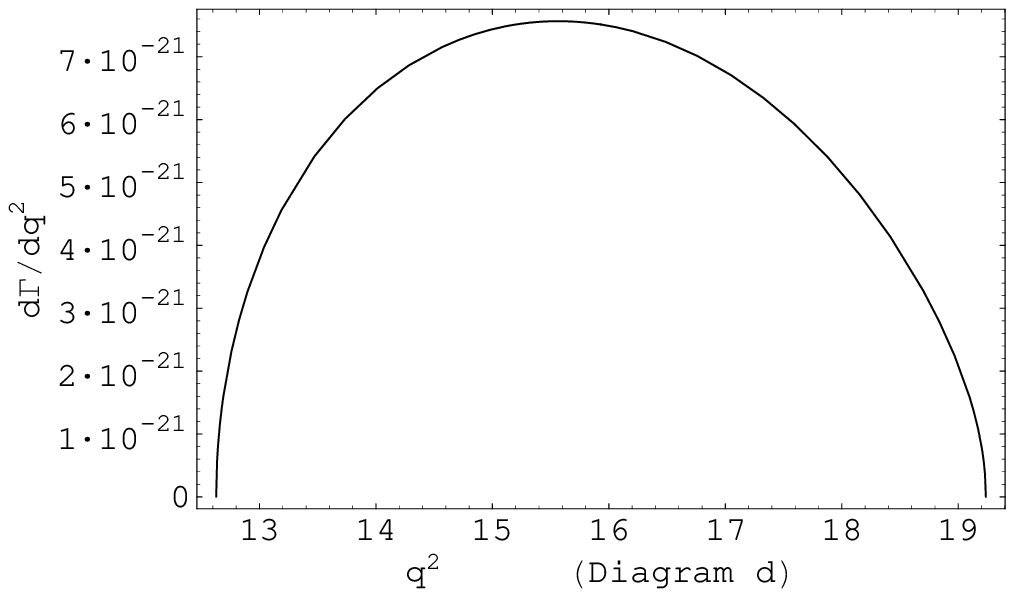}
 \includegraphics[scale=0.6]{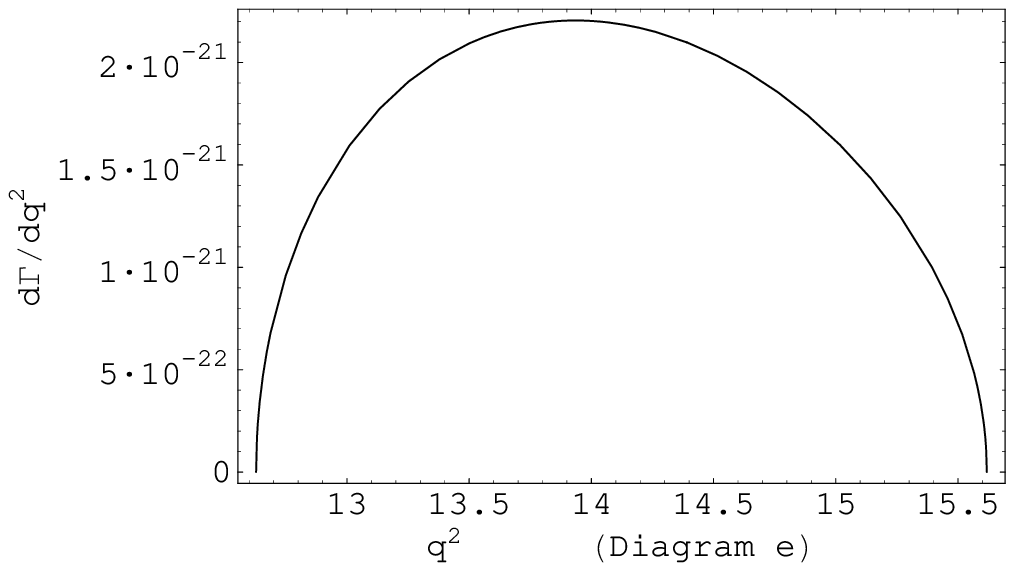}
 \includegraphics[scale=0.6]{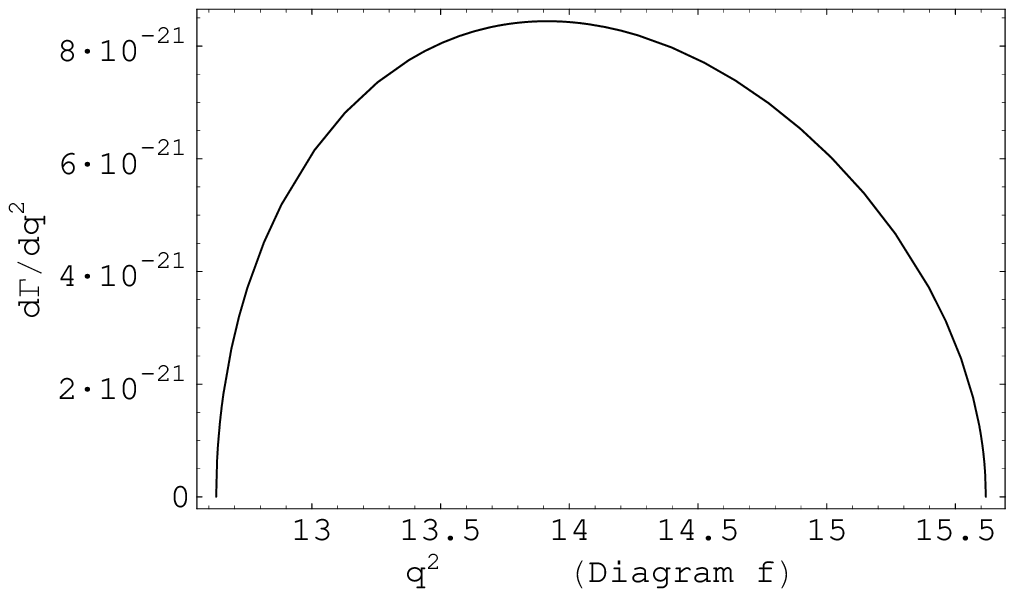}
 \caption{Partial decay widths of the semileptonic  $B\to S\tau^+\tau^-$ decays as functions of $q^2$.
 Diagram a: $B^-\to\kappa^-\tau^+\tau^-$ in scenario 1;
 Diagram b: $B^-\to K^{*-}_0(1430)\tau^+\tau^-$ in scenario 1;
 Diagram c: $B^-\to K^{*-}_0(1430)\tau^+\tau^-$ in scenario 2;
 Diagram d: $\bar B_s^0\to f_0(980)\tau^+\tau^-$ in scenario 1;
 Diagram e: $\bar B_s^0\to f_0(1500)\tau^+\tau^-$ in scenario 1;
 Diagram f: $\bar B_s^0\to f_0(1500)\tau^+\tau^-$ in scenario 2;}
 \label{fig:gamma4}
 \end{center}
 \end{figure}

With the form factors at hand, one can directly obtain the partial
decay width through Eq.~\eqref{eq:pwel} and Eq.~\eqref{eq:pwee}.
Since masses of electrons and muons are very small compared with
$q^2$ in most kinematic region of the semileptonic decays, they will
not produce large effects and are neglected in this work.  In Fig.
\ref{fig:gamma1} and \ref{fig:gamma2}, we give our predictions on
the partial decay width of  $B_{(s)}\to S l^-\bar \nu_l$ ($l=e,\mu$)
and $B_{(s)}\to S \tau^-\bar \nu_l$, respectively.  The diagrams in
Fig. \ref{fig:gamma3} and Fig. \ref{fig:gamma4} are similar but for
the $B_{(s)}\to S l^+l^-$ ($l=e,\mu$) and $B_{(s)}\to S
\tau^+\tau^-$ decays. In Fig. \ref{fig:gamma3}, there exists a small
discontinuity in each diagram, which is caused by the
discontinuities in functions $h(\hat{m}_c,\hat{s})$ and
$h(1,\hat{s})$ in Eqs. (\ref{eq:ypert}). When $l=\tau$ in Fig.
\ref{fig:gamma4}, the discontinuities in the diagrams disappear,
because the origins of $q^2$ axes become $4m_{\tau}^2$ which is
large enough to ensure that the variation of $q^2$ does not pass the
discontinuities in the $h(\hat{m}_c,\hat{s})$ and $h(1,\hat{s})$
functions.

%%----------------------------------------------------------
%\subsection{Branching Ratios}
%----------------------------------------------------------
%%%%%%%%%%%%%%%%%%%%%%%%%%%%%%%%%%%%%%%%%%%%%%%%%%%%%%%%%%%%%%%%%%%%%%%%%%%%%%%%%%%%%%%%%%%%%%%%%%%%
%%%%%%%%%%%     Results of Branch ratios in scenario 1(lnu) %%%%%%%%%%%%%%%%%%%%%%%%%%%%%%%%%%%%%%%%
%%%%%%%%%%%%%%%%%%%%%%%%%%%%%%%%%%%%%%%%%%%%%%%%%%%%%%%%%%%%%%%%%%%%%%%%%%%%%%%%%%%%%%%%%%%%%%%%%%%%
 \begin{table}
 \caption{The total branching ratios for the $b\to ul\bar \nu_l$ in scenario
 1(Unit:$10^{-4}$). The errors are estimated with errors from the
 form factors.}
 \label{tab:branchratios1}
 \begin{center}
 \begin{tabular}{c|c c}
 \hline\hline
 \ \ \                     &$B\to Se\bar \nu_e(\;\mu\bar \nu_{\mu})$    &$B\to S\tau\bar \nu_{\tau}$ \\
 \hline
 \ \ \ $B^-\to\sigma$                     &$0.81_{-0.31}^{+0.52}$    &$0.51_{-0.19}^{+0.33}$   \\
 \ \ \ $\bar B^0\to a_0^+(980)$           &$1.84_{-0.73}^{+1.09}$    &$1.01_{-0.40}^{+0.61}$  \\
 \ \ \ $B^-\to f_0(1370)$                 &$0.29_{-0.13}^{+0.19}$    &$0.13_{-0.06}^{+0.09}$   \\
 \ \ \ $\bar B^0\to a_0^+(1450)$          &$0.67_{-0.29}^{+0.41}$    &$0.28_{-0.12}^{+0.17}$  \\
 \hline
 \ \ \ $\bar B_s\to \kappa^+(800)$        &$1.42_{-0.53}^{+0.82}$    &$0.88_{-0.33}^{+0.52}$  \\
 \ \ \ $\bar B_s\to K^{*+}_0(1430)$       &$0.77_{-0.27}^{+0.37}$    &$0.35_{-0.12}^{+0.17}$  \\
 \hline\hline
 \end{tabular}
 \end{center}
 \end{table}

 %%%%%%%%%%%%%%%%%%%%%%%%%%%%%%%%%%%%%%%%%%%%%%%%%%%%%%%%%%%%%%%%%%%%%%%%%%%%%%%%%%%%%%%%%%%%%%%%%%%%
%%%%%%%%%%%     Results of Branch ratios in scenario 2 (lnu)%%%%%%%%%%%%%%%%%%%%%%%%%%%%%%%%%%%%%%%%%
%%%%%%%%%%%%%%%%%%%%%%%%%%%%%%%%%%%%%%%%%%%%%%%%%%%%%%%%%%%%%%%%%%%%%%%%%%%%%%%%%%%%%%%%%%%%%%%%%%%%
 \begin{table}
 \caption{Same as Table \ref{tab:branchratios1} except in scenario 2.}
 \label{tab:branchratios2}
 \begin{center}
 \begin{tabular}{c|c c}
 \hline\hline
 \ \ \        &$B^-\to f_0(1370)e\bar \nu_e(\;\mu\bar \nu_{\mu})$    &$B^-\to f_0(1370)\tau\bar \nu_{\tau}$ \\
 \hline
 \ \ \ This work   &$1.55_{-0.65}^{+1.53}$ &$0.67_{-0.29}^{+0.68}$  \\
 \hline\hline
 \ \ \        &$\bar B^0\to a_0^+(1450)e\bar \nu_e(\;\mu\bar \nu_{\mu})$    &$\bar B^0\to a_0^+(1450)\tau\bar \nu_{\tau}$ \\
 \hline
 \ \ \ This work   &$3.25_{-1.36}^{+2.36}$      &$1.32_{-0.57}^{+0.97}$\\
 \ \ \ LCSR\cite{YMwang}        &$1.8^{+0.9}_{-0.7}$       &$0.63^{+0.34}_{-0.25}$ \\
 \hline\hline
 \ \ \       &$\bar B_s\to K^{*+}_0(1430)e\bar \nu_e(\;\mu\bar \nu_{\mu})$    &$\bar B_s\to K^{*+}_0(1430)\tau\bar \nu_{\tau}$\\
 \hline
 \ \ \ This work   &$2.45_{-1.05}^{+1.77}$     &$1.09_{-0.47}^{+0.82}$  \\
 \ \ \ LCSR\cite{YMwang}        &$1.3^{+1.3}_{-0.4}$       &$0.52^{+0.57}_{-0.18}$  \\
 \ \ \ QCDSR\cite{MZyang}       &$0.36^{+0.38}_{-0.24}$ &$$\\
 \hline\hline
 \end{tabular}
 \end{center}
 \end{table}
%%%%%%%%%%%%%%%%%%%%%%%%%%%%%%%%%%%%%%%%%%%%%%%%%%%%%%%%%%%%%%%%%%%%%%%%%%%%%%%%%%%%%%%%%%%%%%%%%%%%
%%%%%%%%%%%     Results of Branch ratios in scenario 1(l+l-) %%%%%%%%%%%%%%%%%%%%%%%%%%%%%%%%%%%%%%%%
%%%%%%%%%%%%%%%%%%%%%%%%%%%%%%%%%%%%%%%%%%%%%%%%%%%%%%%%%%%%%%%%%%%%%%%%%%%%%%%%%%%%%%%%%%%%%%%%%%%%
 \begin{table}
 \caption{The total branching ratios for the $b\to sl^+l^-$ in scenario
 1(Unit:$10^{-7}$) with the same error sources as Table
 \ref{tab:branchratios1} and \ref{tab:branchratios2}.}
 \label{tab:branchratios3}
 \begin{center}
 \begin{tabular}{c|c c}
 \hline\hline
 \ \ \                     &$B\to Se^+e^-(\;\mu^+\mu^-)$    &$B\to S\tau^+\tau^-$ \\
 \hline
 \ \ \ $B^-\to\kappa^-$                   &$4.38_{-1.84}^{+2.73}$    &$0.56_{-0.25}^{+0.36}$   \\
 \ \ \ $B^-\to K_0^{*-}(1430)$            &$3.13_{-1.21}^{+1.73}$    &$2.00_{-0.77}^{+1.16}\times 10^{-2}$  \\
% \ \ \ $B^-\to a_0(1450)^-$                 &$$    &$$   \\
% \ \ \ $\bar B^0\to \sigma$               &$$    &$$  \\
 \hline
% \ \ \ $\bar B^0_s\to \kappa^0$          &$$    &$$  \\
% \ \ \ $\bar B^0_s\to K_0^0(1430)$       &$$    &$$  \\
 \ \ \ $\bar B^0_s\to f_0^0(980)$          &$5.21_{-2.06}^{+3.23}$    &$0.38_{-0.16}^{+0.25}$  \\
 \ \ \ $\bar B^0_s\to f_0^0(1500)$         &$1.74_{-0.94}^{+1.14}$    &$2.21_{-1.21}^{+1.32}\times 10^{-2}$  \\
 \hline\hline
 \end{tabular}
 \end{center}
 \end{table}

 %%%%%%%%%%%%%%%%%%%%%%%%%%%%%%%%%%%%%%%%%%%%%%%%%%%%%%%%%%%%%%%%%%%%%%%%%%%%%%%%%%%%%%%%%%%%%%%%%%%%
%%%%%%%%%%%     Results of Branch ratios in scenario 2 (l+l-)%%%%%%%%%%%%%%%%%%%%%%%%%%%%%%%%%%%%%%%%%
%%%%%%%%%%%%%%%%%%%%%%%%%%%%%%%%%%%%%%%%%%%%%%%%%%%%%%%%%%%%%%%%%%%%%%%%%%%%%%%%%%%%%%%%%%%%%%%%%%%%
 \begin{table}
 \caption{Same as Table \ref{tab:branchratios3} except in scenario 2.}
 \label{tab:branchratios4}
 \begin{center}
 \begin{tabular}{c|c c}
 \hline\hline
 \ \ \        &$B^-\to K^{*-}_0(1430)e^+e^-(\;\mu^+\mu^-)$    &$B^-\to K^{*-}_0(1430)\tau^+ \tau^-$ \\
 \hline
 \ \ \ This work   &$9.78_{-4.40}^{+7.66}$ &$6.29_{-2.95}^{+5.71}\times 10^{-2}$  \\
 \ \ \ LCSR\cite{YMwang}        &$5.7_{-2.4}^{+3.4}$       &$9.8_{-5.5}^{+12.4}\times 10^{-2}$ \\
 \ \ \ LFQM\cite{results:LFQM}                  &$1.63$     &$2.86\times 10^{-2}$ \\
 \ \ \ QCDSR\cite{results:QCDSR}                 &$2.09-2.68$        &$(1.70-2.20)\times 10^{-2}$ \\
 \hline\hline
 \ \ \       &$\bar B^0_s\to f^{0}_0(1500)e^+e^-(\;\mu^+\mu^-)$    &$\bar B_s^0\to f^{0}_0(1500)\tau^+ \tau^-$\\
 \hline
 \ \ \ This work   &$10.0_{-3.8}^{+8.5}$     &$0.13_{-0.06}^{+0.12}$  \\
 \ \ \ LCSR\cite{YMwang}   &$5.3_{-1.8}^{+2.3}$       &$0.12_{-0.05}^{+0.08}$  \\
 \hline\hline
 \end{tabular}
 \end{center}
 \end{table}

%%%%%%%%%%%%%%%%%%%%%%%%%%%%%%%%%%%%%%%%%%%%%%%%%%%%%%%%%%%%%%%%%%%%%%%%%%%%%%%%%%%%%%%%%%%%%%%%%%%%%%%

The results for the total branching ratios are collected in Table
\ref{tab:branchratios1}, \ref{tab:branchratios2},
\ref{tab:branchratios3} and \ref{tab:branchratios4}, with the errors
estimated with the errors of the form factors. One can find that the
branching ratios with $\tau$ lepton(s) in the final state are
smaller than the ones without $\tau$ lepton(s), because the large
mass of $\tau$ lepton(s) makes the phase space much smaller. In
Table \ref{tab:branchratios1}, $\frac{Br(B_{(s)}\to
Se\bar\nu_e)}{Br(B_{(s)}\to S\tau\bar\nu_{\tau})}$ is smaller than
two when the scalar meson belongs to the light nonet. While for the
heavy nonet mesons, the value of this ratio is larger than two. The
reason is that more energy is released when the final state is a
light meson, and thus the effect of $m_{\tau}$ on the phase space is
not so evident. In Table \ref{tab:branchratios2} and
\ref{tab:branchratios4}, we also list the predictions in light-cone
sum rules(LCSR) and QCD sum rules(QCDSR), which are smaller than our
predictions. The reason is that we have bigger form factors. Taking
$\bar B^0\to a_0^+(1450)e^-\bar\nu_e$ as an example, the form
factors that contribute are $F_0(q^2)$ and $F_1(q^2)$, with the
relationship $F_0(0)=F_1(0)$. $F_0(0)$ for $\bar B^0\to a_0^+(1450)$
in scenario 2 in this paper is $0.68_{-0.15}^{+0.19}$, while the
corresponding value in \cite{YMwang} is $0.52\pm 0.10$. As a rough
estimation, supposing that corresponding form factors in these two
papers have analogical evolution with respect to $q^2$, the
branching ratio in this paper should be $(0.69/0.52)^2\approx1.7$
times larger.

\section{Conclusions}

In this work, we have studied the $B\to S$ form factors in the PQCD
approach under two different scenarios for the scalar mesons. In
scenario 1, both of the light and heavy nonet are described as the
$\bar qq$ state while in scenario 2, we have only studied the heavy
nonet.  Due to the large decay constant $\bar f_S$, we have found
that most of our predictions are larger than those for the $B\to P$
transition form factors, especially in scenario 2. Contributions
from various LCDAs are explicitly specified. Due to the large masses
of $a_0(1450),K_0^*(1430),f_0(1500)$, their twist-3 LCDAs have
provided more than one half contributions to the form factors in
both scenarios. In scenario 1, the two Gegenbauer moments $B_1,B_3$
for the twist-2 LCDAs have different signs and they give destructive
contributions to the form factors; while in scenario 2, although the
two Gegenbauer moments are small in magnitudes, they give
constructive contributions and induce larger form factors.
Contributions from terms with Gegenbauer moments in the twist-3
LCDAs are also investigated, and we find that these terms do not
give large changes. We also study the semileptonic $B\to Sl\bar\nu$
and $B\to Sl^+l^-$ decays, including the partial decay width and the
integrated branching fractions. Branching ratios of the semileptonic
$B\to Sl\bar\nu$ decays are found to have the order of $10^{-4}$,
while branching fractions of the $B\to Sl^+l^-$ decays have the
order of $10^{-7}$. Compared with results in the previous studies,
our predictions are a bit larger which is caused by larger form
factors. These predictions will be tested by the future experiments.

\section*{Acknowledgements}

This work is partly supported by the National Natural Science
Foundation of China under Grant Numbers 10735080, 10625525, and
10525523. We would like to thank Yu-Ming Wang for fruitful
discussions.

 \appendix
 \section{pQCD functions}
In this part, we collect the  functions which are essential in the
PQCD calculation.

 \label{Appendix:pQCDfunctions}
 \begin{eqnarray}
 t_e^1=\max(t_c\sqrt{(1-x_2)\eta}m_B,1/b_1,1/b_2),t_e^2=\max(t_c\sqrt{x_1\eta}m_B,1/b_1,1/b_2),\label{eq:scales}
 \end{eqnarray}
with $t_c=1$ for the calculation of the central values and
$t_c=0.75$-$1.25$ for error estimation.
 \begin{eqnarray}
 h_e(x_1,x_2,b_1,b_2)&=&K_0(\sqrt{x_1x_2}m_Bb_1)
 \bigg[\theta(b_1-b_2)K_0(\sqrt{x_2} m_B b_1)I_0(\sqrt{x_2}m_B b_2)\nonumber\\
 &&+\theta(b_2-b_1)K_0(\sqrt{x_2} m_B b_2)I_0(\sqrt{x_2}m_B
 b_1)\bigg].
 \end{eqnarray}

\begin{eqnarray}
S_t(x)=\frac{2^{1+2c}\Gamma(3/2+c)}{\sqrt{\pi}\Gamma(1+c)}
[x(1-x)]^c\;, \label{str}
\end{eqnarray}
with $c=0.4$. The Sudakov factor in
Eqs.(\ref{eq:f0})-Eqs.(\ref{eq:fT}) is given by
 \begin{eqnarray}
 S_{ab}(t)=S_B(t)+S_S(t),
 \end{eqnarray}
where
\begin{eqnarray}
S_B(t)&=&s\left(x_1\frac{m_{B}}{\sqrt
2},b_1\right)+\frac{5}{3}\int^t_{1/b_1}\frac{d\bar \mu}{\bar
\mu}\gamma_q(\alpha_s(\bar \mu)),\\
S_S(t)&=&s\left(x_2\frac{m_{B}}{\sqrt
2},b_2\right)+s\left((1-x_2)\frac{m_{B}}{\sqrt
2},b_2\right)+2\int^t_{1/b_2}\frac{d\bar \mu}{\bar
\mu}\gamma_q(\alpha_s(\bar \mu)),
\end{eqnarray}
with the quark anomalous dimension $\gamma_q=-\alpha_s/\pi$. The
explicit form for the  function $s(Q,b)$ is:
\begin{eqnarray}
s(Q,b)&=&~~\frac{A^{(1)}}{2\beta_{1}}\hat{q}\ln\left(\frac{\hat{q}}
{\hat{b}}\right)-
\frac{A^{(1)}}{2\beta_{1}}\left(\hat{q}-\hat{b}\right)+
\frac{A^{(2)}}{4\beta_{1}^{2}}\left(\frac{\hat{q}}{\hat{b}}-1\right)
%\nonumber \\
-\left[\frac{A^{(2)}}{4\beta_{1}^{2}}-\frac{A^{(1)}}{4\beta_{1}}
\ln\left(\frac{e^{2\gamma_E-1}}{2}\right)\right]
\ln\left(\frac{\hat{q}}{\hat{b}}\right)
\nonumber \\
&&+\frac{A^{(1)}\beta_{2}}{4\beta_{1}^{3}}\hat{q}\left[
\frac{\ln(2\hat{q})+1}{\hat{q}}-\frac{\ln(2\hat{b})+1}{\hat{b}}\right]
+\frac{A^{(1)}\beta_{2}}{8\beta_{1}^{3}}\left[
\ln^{2}(2\hat{q})-\ln^{2}(2\hat{b})\right],
%\nonumber \\
%&&+\frac{A^{(1)}\beta_{2}}{8\beta_{1}^{3}}
%\ln\left(\frac{e^{2\gamma_E-1}}{2}\right)\left[
%\frac{\ln(2\hat{q})+1}{\hat{q}}-\frac{\ln(2\hat{b})+1}{\hat{b}}\right]
%-\frac{A^{(2)}\beta_{2}}{16\beta_{1}^{4}}\left[
%\frac{2\ln(2\hat{q})+3}{\hat{q}}-\frac{2\ln(2\hat{b})+3}{\hat{b}}\right]
%\nonumber \\
%& &-\frac{A^{(2)}\beta_{2}}{16\beta_{1}^{4}}
%\frac{\hat{q}-\hat{b}}{\hat{b}^2}\left[2\ln(2\hat{b})+1\right]
%+\frac{A^{(2)}\beta_{2}^2}{432\beta_{1}^{6}}
%\frac{\hat{q}-\hat{b}}{\hat{b}^3}
%\left[9\ln^2(2\hat{b})+6\ln(2\hat{b})+2\right]
%\nonumber \\
%&& +\frac{A^{(2)}\beta_{2}^2}{1728\beta_{1}^{6}}\left[
%\frac{18\ln^2(2\hat{q})+30\ln(2\hat{q})+19}{\hat{q}^2}
%-\frac{18\ln^2(2\hat{b})+30\ln(2\hat{b})+19}{\hat{b}^2}\right],
\end{eqnarray} where the variables are defined by
\begin{eqnarray}
\hat q\equiv \mbox{ln}[Q/(\sqrt 2\Lambda)],~~~ \hat b\equiv
\mbox{ln}[1/(b\Lambda)], \end{eqnarray} and the coefficients
$A^{(i)}$ and $\beta_i$ are \begin{eqnarray}
\beta_1=\frac{33-2n_f}{12},~~\beta_2=\frac{153-19n_f}{24},\nonumber\\
A^{(1)}=\frac{4}{3},~~A^{(2)}=\frac{67}{9}
-\frac{\pi^2}{3}-\frac{10}{27}n_f+\frac{8}{3}\beta_1\mbox{ln}(\frac{1}{2}e^{\gamma_E}),
\end{eqnarray}
$n_f$ is the number of the quark flavors and $\gamma_E$ is the Euler
constant. We will use the one-loop running coupling constant, i.e.
we pick up only the four terms in the first line of the expression
for the function $s(Q,b)$.

 %%%%%%%%%%%%%%%%%%%%%%%%%%%%%%%%%%%%%%%%%%%%%%%%%%%%%%%%%%%%%%%%%%%%%%%%%%%%%%%%%%%%%%%%%%%%%%%%%%%

\end{document}